\documentstyle[11pt,aaspp4,epsf]{article}

\def\plotone#1{\epsfysize=8cm\epsffile[-182 144 392 718]{#1}}
\def\pn{\par\noindent}

\def\gmc{g cm$^{-3}$}
\def\kms{km s$^{-1}$}
\def\msun{M_\odot}
\def\Msun{$\msun$}
\def\Rsun{$R_\odot$}
\def\fefsx{$^{56}$Fe}
\def\cofsx{$^{56}$Co}
\def\nifsx{$^{56}$Ni}
\def\e#1{$\times 10^{#1}$ }
\def\ee#1{$10^{#1}$ }
\def\lum#1{$L_{\rm #1}$}
\def\mas#1{$M_{\rm #1}$ }
\def\etal{et al. }
\def\ltsima{$\; \buildrel < \over \sim \;$}
\def\ltsim{\lower.5ex\hbox{\ltsima}}
\def\gtsima{$\; \buildrel > \over \sim \;$}
\def\gtsim{\lower.5ex\hbox{\gtsima}}
\def\Msunyr{~M_\odot~{\rm yr}^{-1}}

\slugcomment{Accepted 10 November, 1999; To appear in ApJ, Vol. 532}

\begin{document}

\title{Radiation hydrodynamics of SN 1987A:
  \\ I. Global analysis of the light curve for the first 4 months}

\author{Sergei Blinnikov\altaffilmark{1}
\affil{Institute for Theoretical and Experimental Physics,
 117259, Moscow, Russia}}
\author{Peter Lundqvist\altaffilmark{2}
\affil{Stockholm Observatory, SE-133 36 Saltsj\"obaden, Sweden}}
\author{Oleg Bartunov\altaffilmark{1}
\affil{Sternberg Astronomical Institute, 119899, Moscow, Russia}}
\author{Ken'ichi Nomoto\altaffilmark{3} and Koichi Iwamoto\altaffilmark{3}
\affil{$^1$Department of Astronomy and the Research Center for the
Early Universe, University of Tokyo, Tokyo 113-0033, Japan}}
\altaffiltext{1}{E-mail: blinn@sai.msu.su, oleg@sai.msu.su.}
\altaffiltext{2}{E-mail: peter@astro.su.se.}
\altaffiltext{3}{E-mail: nomoto@astron.s.u-tokyo.ac.jp,
                         iwamoto@astron.s.u-tokyo.ac.jp.}

\begin{abstract}

The optical/UV light curves of SN 1987A are analyzed with the
multi-energy group radiation hydrodynamics code {\sc stella}.  The 
calculated
monochromatic and bolometric light curves are compared with
observations shortly after shock breakout, during the early plateau,
through the broad second maximum, and during the earliest phase of the
radioactive tail. We have concentrated on a progenitor model calculated 
by Nomoto \& Hashimoto and Saio, Nomoto, \& Kato, which 
assumes that $14~M_\odot$ of the stellar mass is ejected. Using this model, we
have updated constraints on the explosion energy and the extent of mixing in
the ejecta. In particular, we determine the most likely range of $E/M$ 
(explosion energy over ejecta mass) and $R_0$ (radius of the progenitor). In
general, our best models have energies in the range $E = (1.1 \pm 0.3) \times
10^{51}$~ergs, and the agreement is better than in earlier, flux-limited 
diffusion calculations 
for the same explosion 
energy. Our modeled $B$ and $V$ fluxes compare well with observations, while 
the flux in $U$ undershoots after $\sim 10$~days by a factor of a few, 
presumably due
to NLTE and line transfer effects. We also compare our results with {\it IUE}
observations, and a very good quantitative agreement is found for the first
days, and for one {\it IUE} band ($2500-3000$~\AA) as long as for 3 
months. We point out that 
the $V$ flux estimated by McNaught \& Zoltowski 
should probably be revised to a lower value. 

\end{abstract}

\keywords{supernovae: individual (SN 1987A) --- hydrodynamics --- 
radiative transfer --- shock waves --- stars: interiors --- scattering 
--- supernovae: general}

\section{Introduction}

SN 1987A has provided us with an excellent opportunity to
test the theory of massive star evolution, nucleosynthesis, and
supernova explosion.  The broad band photometric observations, ranging
from ultraviolet to far-infrared, and the resultant bolometric light
curve enable us to probe the physical processes occurring in the
interior of SN 1987A.

The light curve is so sensitive to the hydrodynamics that it is a useful tool 
to infer the progenitor's radius, the distribution of elements, mass of the
ejecta (in particular, the mass of the hydrogen-rich envelope, $M_{\rm env}$),
and the explosion energy, $E$ (e.g., Nomoto \etal 1994 for a review).
Although earlier theoretical models with flux-limited
diffusion approximation are generally in good agreement with
observations (see, e.g., Arnett \etal 1989a, and \cite{hillhoef}
for a review), there are still some uncertainties. % in the theoretical models. 
One of the most important complications is that the supernova atmosphere is
scattering dominated so that the color temperature is much higher than
the effective temperature; the spectrum is a superposition of spectra
emerging from layers with different depths and temperatures (Imshennik \&
Utrobin 1977; Shigeyama \etal 1987;
Pizzochero 1990; H\"oflich \& Wheeler 1999 and references therein).
This effect is crucial to include in order to constrain model parameters.

Previous works on modeling this effect have used supernova atmospheric
codes which rely either on the temperature structure derived from equilibrium
diffusion models (H\"oflich 1991), or on the time-dependent
luminosity from one-group radiation-hydro models (Hauschildt \& Ensman 1994).
To produce light curves, the latter work also used the temperature structure
prescribed
by the one-group radiation-hydro results. In the present work we attempt to
solve the problem of a scattering dominated supernova envelope doing all
calculations time-dependently. We calculate the temperature structure
self-consistently, and we do not make assumptions on the radiative
equilibrium which is strongly violated during the shock breakout.

We analyze the light curve of SN 1987A with a multi-group radiation
hydrodynamics code called {\sc stella} (\cite{blibar93,bli98a}).
The calculated broad band ( $UBV$, {\it IUE} UV)  and
bolometric light curves are compared with observations for the first
$\sim 4$ months after core collapse. From this we
update the constraints on model parameters such as the explosion
energy and the extent of mixing.

We discuss the uncertainties in our predictions due to NLTE effects 
(which are not included in our modeling) by comparing with NLTE atmospheric
calculations done by other authors. We believe that those uncertainties can be
bracketed for the first months of SN 1987A light by two extreme assumptions
on the scattering of photons in spectral lines. It is found that the
predictions for the first hours and days are fairly insensitive to the
assumptions on spectral lines, because of the overwhelming dominance of
electron scattering. So, for those epochs we produce reliable 
predictions for the soft X-ray/extreme UV flash of the supernova.

\section {Radiation Hydrodynamics}

Earlier modeling of the light curve of SN 1987A has adopted the
following types of numerical approaches:

\begin{enumerate}

\item Equilibrium-diffusion radiation hydrodynamics, with a
flux-limiter to ensure a smooth transition from diffusion to
free-streaming regimes (Shigeyama et al. 1987; Arnett 1987, 1988;
\cite{grasimnu87}; Woosley 1988; \cite{viu93}). 
Here {\sl equilibrium} means {\sl
one-temperature} approximation where the radiation and gas have the
same temperature.  Local thermodynamical equilibrium (LTE) is
assumed, and black body spectra are used to obtain monochromatic
broad-band light curves.

\item  Non-equilibrium, one-energy group (gray) radiation hydrodynamics,
where equilibrium between the gas and the radiation is no
longer assumed (i.e., {\sl two-temperature} transfer).  LTE is assumed
and a diluted black body approximation is used for the monochromatic
light curves (Ensman \& Burrows 1992; Mair \etal 1992).

\item  Multi-energy group (non-gray) atmospheric codes combined with
gray radiation hydrodynamics.  Here the atmospheric structures, i.e.,
the distributions of temperature, density, and velocity at $\tau_{\rm sc}
\lesssim 100$ are obtained from the gray hydro code described above (H\"oflich
1991; Hauschildt \& Ensman 1994) and the emerging spectra are computed
taking into account NLTE effects.

\item Full multi-energy group radiation hydrodynamics.
This is the approach in our study, and the code we have
used is called {\sc stella} (Blinnikov \& Bartunov
1993; Blinnikov \etal 1998); a detailed description of new technical
features of the code is given in Sorokina, Blinnikov, \& Nomoto (1999).
LTE for ionization and atomic level populations is assumed.

\end{enumerate}
\pn
{\sc stella} solves the time-dependent equations for the angular moments
of intensity averaged over fixed frequency bands, using up to 300 zones
for the Lagrangean coordinate and up to 100 frequency bins
(i.e., energy groups).  This high number of frequency groups allows one
to have a reasonably accurate representation of non-equilibrium continuum
radiation. There is no need to ascribe any temperature to the radiation:
the photon energy distribution can be quite arbitrary.

The  coupling of multi-group radiative transfer with hydrodynamics
means that we can obtain the color temperature in a self-consistent
calculation, and that no additional estimates of thermalization depth as
in the one-energy group model of Ensman \& Burrows (1992) are needed.
Variable Eddington factors are computed, which fully take into account
scattering and redshifts for each frequency group in each mass zone.
The gamma-ray transfer is calculated using a one-group approximation for
the non-local deposition of the energy of radioactive nuclei.
%rev
Here we follow Swartz, Sutherland, \& Harkness (1995; see also Jeffery 1998),
and we only use a purely absorptive opacity. This should be a good 
approximation.
%rev

In the equation of state, LTE ionizations and recombinations are
taken into account.  The effect of line opacity is treated as an expansion
opacity according to the prescription of Eastman \& Pinto (1993;
%rev
see also Blinnikov et al. 1998).
%rev
Their approach is different from that of Shigeyama \& Nomoto (1990) who
used Rosseland mean opacities for scattering and absorption processes,
where the line opacities were assumed to be 0.009 and 0.01 cm$^2$
g$^{-1}$ for helium and heavier elements, respectively (Los Alamos opacity 
library). Our opacities are also different from those in other 
equilibrium-diffusion or
one-group radiation hydro models, since instead of one single
energy-averaged opacity we need opacities for all our energy groups.

\section{Models}

We have studied two progenitor models in some detail: the evolutionary
model of Nomoto \& Hashimoto (1988) and Saio, Nomoto, \& Kato (1988b)
(see also \cite{shi88,shi90,yamaoka}; Saio, Kato, \& Nomoto 1988a), 
and the non-evolutionary model of
Utrobin (1993). Here we concentrate on the model studied by Nomoto and
co-workers. The results for Utrobin's model (which gives one of the best fits
to the bolometric light of SN 1987A among the equilibrium diffusion models)
are presented elsewhere.

In the evolutionary model, a star with the initial 
mass $M_{\rm ms}$ = 23 $M_{\odot}$ and low 
metallicity $Z$ = 0.005 is evolved from the main-sequence and onwards, 
with the Schwarzschild criterion applied for
convection. During the evolution from blue to red, the stellar mass 
decreases to 16.3 $M_\odot$ and a helium core of \mas{core} = 6.7 $M_\odot$ is
formed.  During the red phase, 0.7 $M_\odot$ of helium is mixed out into the
hydrogen-rich envelope yielding \mas{core} = 6.0 $M_\odot$
and $M_{\rm env}$ = 10.3 $M_\odot$.
The dredge up of the helium enhances the surface helium abundance
to $Y_{\rm surf}$ = 0.43, which is large enough to move the star in 
HR-diagram from the red to the location of Sk~--69$^{\circ}$202 in the blue. 
The resultant luminosity and radius are \lum{0}~= $1.3\times 10^5$
$L_\odot$ and $R_0$~= 48.5 \Rsun, respectively. Because the radius of the
progenitor is
not well constrained in the evolutionary model, we have also constructed
a hydrostatic envelope model with a radius of $R_0$~= 40 \Rsun~
and of $R_0$~= 58 \Rsun~  which are
fitted to the evolved 6 \Msun~ He core.  The 6 \Msun~ He core was
evolved through the iron core collapse (Nomoto \& Hashimoto 1988).
The resultant explosion and explosive nucleosynthesis were calculated as
in Hashimoto, Nomoto, \& Shigeyama (1989) and Thielemann, Hashimoto, \& Nomoto 
(1990). We have assumed in all our models (i.e., irrespective of the explosion
energy) that the mass cut is located at $M_{\rm r} \equiv M_{\rm c}$ = 1.6
\Msun, so that the mass
of the ejecta is $M_{\rm ej}=14.7$ \Msun. Following Shigeyama \& 
Nomoto (1990) we denote our
standard model 14E1. Various suffixes have been added to distinguish
between different models which have different explosion energies and initial
radii, as well as other different physical properties.

For our multi-group computation it is inconvenient to use the
initial evolutionary 14E1 model directly. Instead, the
star was constructed in hydrostatic equilibrium in the same way as
was done in Blinnikov et al. (1998) for SN 1993J:
as we need a much finer zoning in the outer layers in our calculations
than was used in Nomoto \& Hashimoto (1988) and Saio et al. (1988b),
a remap of the original model was done onto another grid. 
%rev
We assume that at the outer boundary (i.e., at $m=M$, $M$ being the total mass
of the star) the material pressure vanishes, $p=0$, and that there is no 
radiation coming from the outside.
%rev
The density structure found in this way is shown in Figure~\ref{nommxfrho}.
%rev
The density of the free blue wind is only $\sim 10^{-16}$ g cm$^{-3}$ when 
scaled as $r^{-2}$ from $10^{17}$ cm and inward to $3 \times 10^{12}$ cm 
(Lundqvist 1999). This falls outside the plot in Figure 1.
%rev

As the shock wave propagates through the star, the interfaces where
the composition changes suddenly from C+O dominated to helium dominated, and
then further out to hydrogen-dominated, are strongly Rayleigh-Taylor
unstable.  This induces mixing of the material before the shock breakout at 
the surface (e.g., \cite{Bandiera84,1989ApJ...344L..65E,arnfm89,BenzTh90},
Hachisu \etal 1990, 1991, \cite{mfa91,basko94}).
Particularly important for the light curve behavior is
the mixing of hydrogen down to the central region. Mixing of $^{56}$Ni 
into the hydrogen-rich envelope also affects the light curve, 
and is decisive for spectral (\cite{viuchuganna95}) and 
X-ray observations. The mixed abundance distribution in Figure~\ref{nommxfchm} 
is inferred from the comparison of the X-ray and $\gamma$-ray light curves 
and spectra with observations (Kumagai et al. 1989).
%rev
The outermost mixed  $^{56}$Ni
corresponds to the velocity $\sim 4000$ /kms in the best fitting model of
the light curve (see Fig.~\ref{14e1vchm} below). 
The mixing is artificial,
since 2D modeling failed to distribute $^{56}$Ni further out than $\sim
2500$ /kms (Hachisu \etal.  1990, 1991). The density profile was only
marginally changed by the mixing.
%rev

We have tested explosion energies in the range $E = (0.7 - 1.5) \times 10^{51}$
ergs and named the runs 14E0.7, 14E1, 14E1.3, ...,  etc., where `14' denotes 
the ejecta mass in solar units, see Table~\ref{runs}.
All models in the runs labeled without the suffix `U' have the mixed
composition shown in Figure~\ref{nommxfchm} already prior
to the model explosion. 
%rev
Each model was exploded by the
deposition of heat energy in a layer of mass $\sim0.03$ \Msun\ outside of
1.6 \Msun. Since {\sc stella} does not include nuclear burning,
preservation of the
same mixed composition in the ejecta is assured.
%rev
The explosion energies given above refers to the
asymptotic kinetic energy of ejecta. The heat energy injected for a
simulated explosion is higher by $\sim 0.7 \times 10^{51}$ ergs, i.e., by the
gravitational binding energy for the presupernova model. A small
fraction ($\sim 7\times 10^{48}$ ergs)
goes to the photon energy  which is radiated away.  For other models, 14E1U,
and 14E1.2U, the unmixed composition in Figure~\ref{unmxchm} was used, and in 
14E1A, we treated the opacity as pure absorption for the same total extinction
(which, however, is often dominated by scattering in reality). For the latter
model the entry for ``forced $\chi_{\rm abs}$'' is ``yes'', 
i.e., $\chi_{\rm abs}$
was artificially set equal to the total extinction. The suffix `H' denotes
the model having the same density structure as the standard one, 14E1, but
the abundance of hydrogen in the outer layers is enhanced to the solar value.
The suffix `R' is added for the models with non-standard initial radius
(see Table~\ref{runs}). Some other models from the  Table~\ref{runs} are
discussed below.

\placetable{runs}

\section{Hydrodynamics and Shock Breakout}

The shock wave arrives at the surface of the star at time $t_{\rm prop}$,
which for different values of the initial radius $R_0$, the ejected
mass $M_{\rm ej}$, and explosion energy $E$, could be approximated by:

\begin{equation}
 t_{\rm prop} \approx 1.6 \biggl({R_0\over 50 R_\odot}\biggr)
 \Biggl[\biggl({M_{\rm ej}\over 10M_\odot}\biggr)
 /\biggl({E\over 1\times 10^{51}
 {\rm erg}}\biggr)\Biggr]^{1\over 2} {\rm hours} \; .
\label{tprop}
\end{equation}

This expression is consistent with Shigeyama \etal (1987). 
We compare the estimate (\ref{tprop}) with computations in Figure~\ref{tmre}.
The change in the velocity profile in Figure~\ref{nomv} shows how
the materials are accelerated near the shock breakout until they reach
homologous expansion. The maximum velocity at the outer edge 
is $\sim 32,500$ \kms\ for 14E1 (which is somewhat lower than for 14E1.3 
in Figure~\ref{nomv}). Further acceleration is limited by the 
inefficiency of the radiative precursor of the shock 
(see a semi-analytic approach reviewed by Nadyozhin 1994; see also
Imshennik \& Nadyozhin 1988). We will discuss the influence of various
assumptions on the maximum velocity and other details of shock breakout
elsewhere.
Here we just note that velocities of the order $\sim (3-4)\times 10^4$ \kms~are
in very good agreement with the value found from the absorption feature 
of Mg~II~$\lambda$2800 in the early {\it IUE} spectra (\cite{pun95}).
This has important implications for our understanding of the density structure
of the circumstellar medium of the supernova (Chevalier 1999; Lundqvist 1999).
More specifically, it indicates that the density of the blue supergiant wind
must have been very low, corresponding to a mass loss rate of
only $\lesssim 10^{-8} \Msunyr$.

The density distribution in the outer part is well approximated by a
power law $r^{-8.6}$ as found in Shigeyama \& Nomoto (1990), but the very
outermost layers are much steeper (see Fig.~\ref{rholgr}). In between there is
a dense shell, which was also found in non-equilibrium radiation hydrodynamic
modeling (\cite{bli91,bnb91,ens92}), but missed in the equilibrium diffusion
modeling. Note that the density in the central parts computed by {\sc stella}
is much smoother than shown in Figure~6 by Shigeyama \& Nomoto (1990).

\section{Early Light Curve}

After the shock breakout, the early light curve up to $t\sim$ 25 days
is powered by the diffusive release of the internal energy of the
radiation field that is established by the shock wave.
The bolometric light curve reaches its maximum luminosity
of $L_{\rm bol} \sim$ (4--9) $\times 10^{44}$ ergs s$^{-1}$
(Table \ref{earlight} and Fig.~\ref{nommxftphea}) immediately after shock
breakout, and then drops rapidly by many orders of magnitude 
in $\sim 10$~days.

\placetable{earlight}

The total energy radiated during the first two days amounts to $\sim$ \ee{47}
ergs (Table \ref{earlight}), but most of the radiation is emitted in a
soft X-ray/EUV burst, and was not observed. However, the burst had the
important effect that it ionized the surrounding gas (e.g., Lundqvist \&
Fransson 1996; Sonneborn et al. 1997; Lundqvist 1999). The resultant ionization
of the circumstellar material is commented on briefly in \S 7.1, and will be
compared with the observations in greater detail in Lundqvist, Blinnikov, \&
Bartunov (1999a).

After the burst, the ejecta expand so rapidly that the interior
temperature (both of the matter and radiation) decreases almost
adiabatically as $r^{-1}$. As a result, the bolometric luminosity decreases 
sharply to $L_{\rm bol} \sim$ (2 -- 3) \e{41} ergs s$^{-1}$ to form a minimum 
of the bolometric light curve. Figure~\ref{14elum3} demonstrates that the model
with $E = 10^{51}$~ergs (i.e, 14E1) gives the best agreement with the
observed bolometric flux. Note that the agreement for 14E1 is much better
than in Figure~7 of \cite{shi90}). (Note also the higher resolution in our
figure than in the figure of \cite{shi90}). The light curves computed by the
flux-limited diffusion in Shigeyama \& Nomoto (1990) produce a short plateau, 
$\sim$ 20 days (see Figs.~16 -- 19 in \cite{shi90}). This is neither seen
in the observations, nor in our models. The light curves computed here by
the full radiative transport are in better agreement with observations 
for {\em the same models} (cf. Fig.~\ref{14elum3}) 
%rev
suggesting that the more accurate method of modeling gives results which
are closer to reality.
%rev

The luminosity at this phase is lower than for typical Type II-P supernovae
by a factor of 10 -- 20. This is due to the small initial radius which
leads to a low luminosity because a much larger fraction of the radiation
field energy is lost by $P$d$V$ work than in ordinary Type II-P supernovae. 
This is well-known from early modeling of low-luminosity Type II-P supernovae
(\cite{imshnad64}; Chevalier 1976). For SN 1987A, the low luminosity was
successfully demonstrated in the models of Shigeyama \etal (1987),
Arnett (1987), Grasberg \etal (1987), Woosley (1988), Woosley, Pinto, \&
Eastman (1988) and Utrobin (1993).

%rev
In Figure~\ref{nommxftphea}, we show the changes in the {\sl color}
temperature, $T_{\rm c}$, of the best blackbody fit to the flux, along
with the effective temperature, $T_{\rm eff}$, defined by the luminosity and 
the radius of last scattering $R$ through $L=4\pi\sigma T^4_{\rm eff}R^2$
(see Blinnikov \etal, 1998, for details of finding $R$ and from that 
$T_{\rm eff}$).
%rev
The maximum value of $T_{\rm c}$ is $\sim 1.2 \times 10^6$~K for model 14E1
(see Table \ref{earlight}), which is higher than the $7.6 \times 10^5$ K 
in H\"oflich's (1991) NLTE time-dependent calculation for 14E1.25, 
but similar to the temperatures found by Ensman \& Burrows (1992).
Our results are in very good agreement with the estimates of Imshennik \&
Nadyozhin (1988, 1989). We emphasize that our multi-group radiative transfer
with hydrodynamics obtains this temperature in a self-consistent way,
and no additional estimates of the thermalization depth (like in the one-group
model of Ensman \& Burrows 1992) are needed. The large difference between
color and effective temperatures is due to scattering
(e.g., Sobolev 1980; Kolesov \& Sobolev 1982; H\"oflich 1991;
Wagoner \& Montes 1993); the average energy of the photons is higher
than that corresponding to the value of $T_{\rm eff}$. The effect is a
deficit of photons in the visual at maximum light compared to a model
with forced absorption (Fig.~\ref{max1st}). This effect continues throughout
the first day after shock breakout.
In Figure~\ref{scaabs} we demonstrate the dominance of scattering in
extinction for high temperatures. 

Figure~\ref{14etph3} shows $T_{\rm c}$ for the same three runs
as in Figure~\ref{14elum3}, and one can see that $T_{\rm c}$ is rather
insensitive to $E$ (except for the maximum $T_{\rm c}$ at shock breakout, 
as displayed in Table~\ref{earlight}).

During the first day, the visual luminosity increases because the
intensity peak is rapidly shifted into the optical due to the decreasing
photospheric temperature.  In order for the optical flare-up
of the supernova to be seen at 6 mag at $t=3$ h, the condition $t_{\rm
prop}<3$ h (Eqn. [1]) should be satisfied, which requires a relatively
large $E/M_{\rm env}$ and small $R_0$ (Shigeyama \etal 1987; Woosley et al. 
1987). Also, the ejecta and the radiation field should have expanded rapidly
so that the radiation temperature becomes lower and the radius of the
photosphere becomes larger sufficiently fast.  Therefore, the expansion
velocity, and thus $E/M_{\rm env}$, should be larger than certain
values for a given initial radius.

Figure~\ref{vearly} shows the $V$ light curve for the model 14E1.21 with
realistic scattering dominated opacity for the first hours of SN 1987A.
Here, and for other models below, we have used a distance modulus of 18.5.
(See Walker 1999, for a discussion on the uncertainty of this value.)
We note that the $V$ curve exhibits an early local minimum which does not
exist in equilibrium diffusion models (e.g., \cite{woo88,arn88}; a small
minimum was found also by \cite{viu93}).  H\"oflich (1991) ascribed 
%rev
this to a non-LTE effect, and used the location in time of the minimum to
constrain $E$. However, we too recover the local minimum despite our LTE
approach. The reason for the minimum is that when the bolometric flux
continues to fall (Fig.~9) after shock breakout, the bolometric correction
overcomes the effect of the falling bolometric flux. The visual luminosity 
then increases because the intensity peak is rapidly shifted into the 
optical due to the decreasing photospheric temperature. Regardless of the 
exact cause of the difference between our and H\"oflich's results on the one
hand, and equilibrium diffusion models on the other, the first maximum in $V$ 
in our models is just on the level of Jones' limit
%rev
(\cite{wamp87}) even for our 14E1.3 run, where the energy of explosion
is $\sim 9\times 10^{49}$ ergs higher (see Table \ref{runs}) than the model
14E1.25 used by H\"oflich (1991; see also H\"oflich \& Wheeler 1999). The 
earlier appearance of the peak of the $V$-flux in our run is due to this
higher energy.

Compared with observations, and especially the first $V$ observation
by McNaught \& Zoltowski (1987; later revised by West \& McNaught 1992),
the calculated $V$ curve in Figure~\ref{vearly} rises too slowly. In
equilibrium diffusion modeling (e.g., \cite{arn88,woo88,shi90}) the $V$ flux
rises much faster.
The reason for this difference is that equilibrium diffusion models do not
care whether the total extinction is absorptive or due to scattering.
The spectra are assumed in those models to be blackbody
with the temperature equal to $T_{\rm eff}$. This is not a good approximation,
since in reality the thermalization of photons takes place well below the
surface of the last scattering resulting in $T_{\rm c} \gg T_{\rm eff}$.
In our run 14E1A we have assumed purely absorptive extinction, i.e., 
we force even the electron scattering to act as true
absorption so that the thermalization of the photons occurs at an optical 
depth of order unity. We thereby reproduce the results obtained in equilibrium 
diffusion models. As a matter of fact, the $V$ curve in 14E1A rises even 
faster than was observed. (See Fig.~\ref{nommxfea} which shows the $UBV$
light curves for 14E1 and 14E1A for the first 5 days.)
There is of course no physical reason why electron scattering
should be treated as absorptive. From this it is clear that that an accurate
treatment of scattering is crucial to constrain $E$. We will discuss this now.

H\"oflich (1991) was the first to show that non-equilibrium effects
influence the $V$ light curve of SN 1987A drastically; the flux in $V$
is $\sim 2$ magnitudes lower for the first hours than in the equilibrium gray
atmosphere case.
H\"oflich (1991) tried a higher explosion energy than in 14E1 of Shigeyama \&
Nomoto (1990) to compensate for this reduction of flux by adopting the model 
14E1.25 computed in the flux-limited equilibrium diffusion approximation
by Shigeyama \& Nomoto (1990). He then found a marginal agreement with the
first data points of McNaught \& Zoltowski (1987). The same results have since
been presented in H\"oflich \& Wheeler (1999).
Compared with our $V$ light curve of 14E1.21 and 14E1.3 (see Figs.~\ref{vearly} 
and \ref{eashv2}) the rise of the $V$ curve in H\"oflich's (1991) 
14E1.25 model is significantly faster. To clarify this discrepancy, 
one should note that H\"oflich used the temperature
structure of an equilibrium diffusion model, which can be appreciably 
different from that in our full transport models.
We cannot say with certainty that this fully explains why our results
and those of H\"oflich differ so much.
%rev
It could be due to that the expansion opacity should be treated 
differently in the energy equation
than is described in H\"oflich (1990; see 
Blinnikov 1996, 1997). 
%rev

There are, of course, uncertainties also in our models. In particular, the
role of NLTE effects needs to be further examined, but we cannot
envisage that they are able to explain the big difference between us and 
H\"oflich (1991) for the
first hours after shock breakout. At this epoch the extinction is totally
dominated by electron scattering and spectral lines are not so important,
see Figure~\ref{scaabs}.
We see virtually no difference in our results for the first day when we treat
the lines as fully absorptive, or as totally scattering dominated. The NLTE
effects set in later, and are very important after a couple of weeks
(cf. \cite{baron96}). We are therefore confident that our results for
this epoch are more accurate than H\"oflich's (1991).

We note that even H\"oflich's $V$-flux is lower than the observed after the
new reductions by West \& McNaught (1992). A possible explanation to the
discrepancy between models and observations is that the stars, used to
calibrate the early plates of the supernova by West \& McNaught (1992),
are too cool for an object with a color temperature of $T_{\rm c} \sim 10^5$K.
(We find at $t=0.128$ day 
that $T_{\rm c} = 1.14 \times 10^{5},~1.05 \times 10^{5},~9.8\times 10^{4}$~K
for the runs 14E0.7, 14E1, and 14E1.3, respectively.)
The temperature at $t=0.128$ days decreases with increasing explosion energy
because of the earlier emergence of the shock and the faster adiabatic cooling.
It should be emphasized that our models are much hotter than equilibrium
diffusion models (e.g., \cite{arn88,woo88,viu93}), and a comparison between
observations and our results is therefore more sensitive to calibration errors
than are equilibrium diffusion models. As we will see in \S 6.1 our models fit
the early {\it IUE} observations well, and since these observations are less
likely to have the same error, we cannot exclude calibration errors to be the
cause of the mismatch in $V$. {\it This would mean that the early true $V$ flux 
was lower than hitherto believed}. We point out that we have changed various
parameters in our models to try to make our $V$ flux increase faster and 
thereby fit the observations better. These experiments included enhancing the
iron abundance,
and varying the explosion energy and presupernova model within the limits
allowed by the global light curve. However, none of these attempts reduced the
discrepancy. (See, e.g., the results for two different initial radii in 
Fig. \ref{eashv2}.)

\section{Before and After the Peak of the Light Curve}

After the minimum around day $\sim$ 10, the observed bolometric light
curve showed an almost exponential increase up to day $\sim$ 60, and
subsequently formed a plateau-like broad peak around day $\sim$ 100. After
a relatively rapid drop, the luminosity then declined slowly between
$t$ = 120 -- 400 days at the rate of \cofsx-decay. The energy source
responsible for the broad peak of the light curve, and the tail, is therefore
without doubt the radioactive decay 
of \nifsx $~\rightarrow$ \cofsx $~\rightarrow$ \fefsx\ .
The total mass of initial \nifsx\ in our models 
is $M_{\rm Ni} \approx 0.078\msun$, which is the same mass as in the models
of Shigeyama \& Nomoto (1990).
The theoretical bolometric light curves for models 14E1, 14E1M, and 14E1U
with a different extent of mixing are shown in Figure~\ref{nommxfbol}.
It is clearly seen that the shape of the modeled light curve is strongly
dependent on the distribution of hydrogen and \nifsx\ in the ejecta, i.e., the
amount of mixing that has occurred.

For the model with standard mixing (14E1 with $^{56}$Ni mixed out
to $v \sim$ 4000 \kms, Fig.~\ref{14e1vchm}), there is significant heating of
the outer layers due to radioactivity.  The heating becomes noticeable in
the light curve already at $\sim 10$~days, and then forms a smooth increase in
the optical light curve up to the peak, as observed.

For the unmixed model 14E1U (Figs.~\ref{unmxchm}, \ref{14e1uvch}),
the shape of its bolometric light curve starts to differ from that of
14E1 already at $\sim$ 10~days. For the unmixed case, the increase in the
luminosity due to radioactive heating is delayed until $t=35$ days, which 
causes a dip to appear in the light curve around day 30, and makes the 
light curve in the subsequent phase rise faster than in the mixed model. 
These properties are clearly incompatible with the observations.

For the case with intermediate mixing (14E1M), \nifsx\ is mixed
out to $v \sim$ 2500 \kms , while hydrogen is mixed as in the mixed model 14E1
(Fig.~\ref{14e1uvch}). Also in this model the appearance of radioactive heating
occurs too late to be compatible with observations (see Fig.~\ref{nommxfbol}).
This suggests that mixing of \nifsx\ out to $v \sim$ 4000 \kms\ is needed.
This conclusion is independent of the radiation transfer scheme used
(Nomoto, Shigeyama, \& Hashimoto 1987; Woosley 1988; Shigeyama \etal 1988).
The mixing out to large velocities is also supported by the redshifted
feature at $\sim 3900$ \kms\ observed by Haas et al. (1990; see 
also~\cite{viuchuganna95}). We note, however,
that Utrobin (1993) obtains a good bolometric light curve 
without \nifsx\ mixing, and that Kozma \& Fransson (1998a,b) 
do not need to mix nickel out to more than $\sim$ 2000 \kms\ to model iron 
line profiles at late epochs. We postpone a more detailed discussion on
mixing to a future paper.

It is also interesting to compare the results of our non-equilibrium radiative 
transfer modeling by {\sc stella} with the models in Figures~16 -- 19 of 
Shigeyama \& Nomoto (1990). One striking difference, already noted above, is
that the shape of the minimum near day 10 is reproduced much better with our
non-equilibrium modeling.
Another difference
is that the shape of the light curve around maximum (at $\sim 3$ months) is
smoother in models with realistic scattering opacity (see Fig.~\ref{bol3en}
below), than in models with forced absorption
like 14E1A (see Fig.~\ref{nommxfabbol}), and hence also in equilibrium
diffusion models since the latter two types of model are closely
related. These two models also show the same postmaximum sharp decline (cf.
\cite{viu93}), which can be understood in terms of enhanced emission according 
to Kirchoff's law.

We note that the rising part of the light curve is modeled better with
enhanced absorption in spectral lines than with scattering lines (see
Fig.~\ref{nommxfabbol}
for the bolometric luminosity, and Figs.~\ref{14e1ubv}, \ref{14e1subv}
for $UBV$ colors). This can perhaps be explained by NLTE effects. In
particular, the effect of fluorescence could be important
(see \cite{baron96,lietal,li96}, and references therein). This cannot be
studied directly by {\sc stella}, but, as found by Baron et al. (1996),
forced absorption in spectral lines can reproduce some properties of the
fluorescence
%rev
as fluorescence is a form of thermalization;
%rev
spectra of LTE models with absorptive lines are very similar to
full NLTE spectra, while LTE models with scattering lines are far from reality
(see also Eastman 1997; \cite{bli98a}). This hints why the rising part is
modeled well when we apply forced absorption in lines.

Another cause for the deviation of the rising part of the light curve from
what was observed could be that the distribution of hydrogen is different 
from that in our model. This effect was investigated by Utrobin (1993). The
hydrogen distribution affects the light curve because it determines how the 
hydrogen recombination front propagates into the ejecta (e.g., Nadyozhin 1994;
Shigeyama \& Nomoto 1990). Because electron scattering is the main source of
opacity, the opacity decreases sharply in the outward direction at the
recombination front. This causes the photosphere to become associated with the
recombination front.  When the ejecta pass through this front their temperature
quickly decreases to $\sim 5500$ K. This is seen as a change in the temperature
profile in Figures~\ref{settm} and \ref{settr}. One can locate
the photosphere in Figure~\ref{settm} at the point where the radiation
temperature starts to deviate from the material temperature appreciably.  We
note that the recombination front is much broader than that in Shigeyama \&
Nomoto (1990) because of the large contribution of line opacity to the total
opacity in our calculations.  The effects of non-equilibrium transport are
also important for the width of the recombination front.

As seen in Figures~\ref{settm} and \ref{setlm} the photosphere propagates
inward in mass, while the material expands outward. For a certain period, the
recombination front is almost stationary in radius (Figure~\ref{settr}).
If $T_{\rm eff}$ were constant, then this would
result in an almost constant bolometric luminosity, i.e., in a perfect
plateau of the light curve, where the duration of the plateau stage depends 
on how deep into the star hydrogen has been mixed. For a deeper mixing of 
hydrogen, the plateau lasts longer. When the photosphere enters into 
layers which lack sufficient amounts of hydrogen, the hydrogen recombination 
front disappears and the plateau phase is terminated. This is the case
for typical Type II-P supernovae, where radioactive heating does not show 
up until the very end of the plateau stage (Eastman et al. 1994).
SN 1987A is quite different in this respect, since radioactivity is important
also at early epochs; the diffusion flux from the radioactive energy release
starts to dominate over the diffusion flux from the recombination of hydrogen
already at $\sim 6$ weeks after the explosion. This is clearly seen in
Figure~\ref{setlm} (cf. Fig.~12 in \cite{shi90}, and Fig.~11 in \cite{viu93}).

The relation between the duration of the plateau phase and the depth
of the hydrogen layer can be given more quantitatively. Suppose that
hydrogen is mixed down to a shell where the expansion velocity of the
%rev
hydrogen-rich layer is $v_{\rm H}$.
Then this velocity is related to the observed quantities for the 
plateau phase as
\begin{equation}
v_{\rm H}\approx 1300~ 
\left( {L_{\rm pl}\over 8.5 \times 10^{41}\hbox{ ergs
s$^{-1}$}}\right)^{1/2}~
\left({t_{\rm pl}\over 100 \hbox{ d}}\right)^{-1} \hbox{ km s$^{-1}$} \;
\label{vH}
\end{equation}
where $t_{\rm pl}$ is the time at the end of the 
plateau, $L_{\rm pl}=4\pi\sigma T_{\rm eff}^4 R^2$ the
luminosity at $t\sim t_{\rm pl}$, and $T_{\rm eff} \sim$ 5500 K because of
the association of the photosphere with the hydrogen recombination front. 
The time $t$
is for the freely expanding ejecta, $t = R/v$.
For example, if we take $L_{\rm pl}=2\times 10^{41}$ ergs s$^{-1}$,
and $t_{\rm pl}=30$ days, as in the unmixed model 14E1U (Fig.~\ref{nommxfbol}),
we find from Eqn.(\ref{vH})  $v_{\rm H}\approx 2000$ \kms\,
%rev
which is in good agreement with Figure~\ref{14e1uvch}.

In the standard model, 14E1, hydrogen is mixed down
to $M_{\rm r} \approx 2 \msun$, i.e., only $\approx 0.4\msun$
outside $M_{\rm c}$. The expansion velocity at that radius is only $\sim$
1000 \kms. If we adopt $t_{\rm pl} \sim 100$ days, Eqn.~\ref{vH}
gives $v_{\rm H} \sim$ 1300 km s$^{-1}$ as is required
from observations. Equation (\ref{vH}) thus appears to give a reasonable
estimate of $v_{\rm H}$ when we put $t_{\rm pl}$ equal to the time
when the second maximum ends. However, we caution that this result should
not be overinterpreted. Equation (2) gives the photospheric velocity as long 
as $T_{\rm eff}$ is 5500 K, which is roughly the case for SN 1987A also for 
the second maximum. But the formula also assumes that the temperature is 
governed by the presence of hydrogen. This is not a unique statement, as the
complicated thermal balance may settle around this temperature also without 
the presence of hydrogen. Even if hydrogen is mixed far into the core, the 
end of the second maximum is likely to give only limited information about the
minimum velocity of hydrogen.

\subsection{Broadband fluxes}

Ideally, both spectra and $UBV$ colors should be obtained by full
NLTE modeling (see arguments put forward by Eastman 1997). For SN 1987A
spectral NLTE modeling has been made by several groups (see, e.g.,
\cite{schmutz,hoefdiss,takeda,mazzlb,dusch,mazzchug,hoefwhee}
and references therein), while NLTE modeling of colors are given in the
literature only for the first days after the explosion
(\cite{eastkir,hausens}).
Our approach cannot add to this NLTE modeling, but as discussed by, e.g.,
H\"oflich (1995, see also references therein), LTE modeling is useful even for
such rapidly expanding objects as Type Ia SNe, and should therefore also give
some insight to the physical conditions of SN 1987A. Inspired by this, we
compare our LTE results with observations in Figure~\ref{14e1ubv}.
The $B$ and $V$ light curves are in surprisingly good agreement with
observations.  Likewise, for the $U$ band the agreement is satisfactory
for the first days, though at later epochs the modeled flux is too high.
Nevertheless, the shape of the light curve qualitatively reproduces the
observations.

There is also a qualitative agreement between our calculated fluxes and
the observed fluxes in {\it IUE} bands (\cite{pun95})
(see Figs.~\ref{iue10},~\ref{iue100}). In particular, the agreement is good
for the first 10 days for the SWP and the two LWP bands, while the modeled flux
overshoots early for the LWP 3000 -- 3300 \AA\ band, like it does for the
modeled $U$ flux. Later, the modeled flux in the bands with the shortest
wavelengths undershoots, while for the range 2500 -- 3000 \AA, the agreement 
is still fairly good. The modeled flux in the band with the longest
wavelengths continues to be high for the first 100 days (Fig.~\ref{iue100}).

The disagreement between modeled and observed UV fluxes
after $\sim 100$~days is not surprising, because the LTE modeling at that epoch
becomes quite unrealistic when there is no longer a true photosphere.  For
earlier epochs, there can be several causes for deviations. The early drop of
the modeled fluxes in the shortest IUE bands could signal that the photosphere
is somewhat hotter than in the 14E1 model; the flux here falls into the Wien
part of the spectrum and it is exponentially sensitive to the temperature. It
is harder to explain why the observed UV flux for $\lambda > 3000$~\AA\ and
the flux in the $U$ band fall below what we predict. Within the LTE approach
this could mean that the expansion opacity is not complete in the Eastman-Pinto
(1993) approximation. The line list of Eastman \& Pinto includes $\sim 10^5$
lines, but 
%rev
they and
%rev 
Baron et al. (1996) have pointed out that it is also necessary to
include millions of weaker lines (see also H\"oflich 1995), mostly of iron.
Should the Eastman-Pinto list be sufficient, then one might think of enhancing
the abundance of iron group elements in the outer layers of the supernova to
increase the opacity. Our experiments with such an enhancement show that we can
bring the $U$ flux in much better agreement with observations.
There could also be other causes for our too strong UV flux, but it seems
reasonable to assume that line opacities are somehow involved. This is
highlighted by the good agreement between modeled and {\it IUE} fluxes in the
2500 -- 3000 \AA\ range where there are relatively few spectral lines.
This could perhaps indicate that the envelope is contaminated with heavy 
elements which could give a higher opacity in the UV.

In this context we note that Pun et al. (1995) do not cite Wagoner,
Perez, \& Vasu (1991) correctly when they state that ``the expansion opacities
in the wavelength region 1000--4000 \AA\ increase by a factor of more than 100
as the temperature of the atmosphere drops from 12,000 to 5000 K''. We point
out that it is not the expansion opacity in Wagoner et al. (1991; see
also \cite{eastpin}) which increases by this number. The correct statement
is that the ratio of expansion opacity to electron scattering increases by a 
factor of $\gtrsim 100$. It does so because the electron scattering opacity
drops drastically due to recombination as the temperature is lowered.

We have not tried to include millions of weak lines in our models to check
whether this can bring the UV observations and our LTE models in better
agreement. If such an experiment would fail, and the metal content of the
envelope is not unusually high (cf. above), then one has to consider NLTE
effects already for a few days after the explosion. The main NLTE effect
here could be the excitation of hydrogen from its second principal level, $n
= 2$, perhaps creating an optically thick Balmer continuum. From NLTE
atmospheric calculations (see, e.g., \cite{schmutz,takeda,dusch}) we know that
early overpopulation of $n = 2$ is present in SN 1987A, though it is not
sufficient to explain the observed absorption in the Balmer range,
unless there is a direct excitation due to radioactivity.
While it is certainly important to include nonthermal excitation in
late spectra hydrogen (e.g., \cite{xuetal}; Kozma \& Fransson 1998b),
the same effect operating at early times could be one more hint of extremely
efficient outward mixing of radioactive material into the hydrogen-rich 
envelope.  The possibility that hydrogen could be excited as a result of
circumstellar interaction seems much more unlikely because of the very low
density inferred for the circumstellar gas (\S 4; Chevalier 1999; Lundqvist 
1999).

\subsection{Parameters of the Photosphere}

We present in Figures~\ref{tpfit} -- \ref{vphot} the comparison of our
numerical results with the ``photospheric'' parameters found by observers.
We put this in quotes since what is given by observers are not really the
parameters of the photosphere, but the best blackbody fit
temperature $T_{\rm obs}$ (which is higher than $T_{\rm eff}$), and the
radius, $R_{\rm obs}$, found from $L=4\pi\sigma T_{\rm obs}^4 R_{\rm obs}^2$.
We emphasize that $R_{\rm obs}$ is substantially smaller than the radius
of the true photosphere, $R_{\rm ph}$, especially at the earliest stages.

\section{Dependence on Model Parameters}

\subsection{Explosion Energy}

In the above discussion, the light curve has mainly been used to probe the
internal abundance distribution. Figure~\ref{bol3en} shows how the bolometric
light curve depends on $E$. For a given ejecta mass and using the 
standard mixing, we can find constraints on $E$ from the light curve
(Shigeyama \etal 1987, 1988; Nomoto \etal 1987, 1994; Woosley 1988;
Woosley \etal 1988; Arnett \& Fu 1989; \cite{ims92}). 
First, the luminosity near the minimum
around day 10 (which corresponds to the early short plateau in the $V$ curve)
is almost proportional to $E$ (Litvinova \& Nadezhin 1990, Popov 1993), 
thus providing an important 
constraint on $E$.  Second, the time of the peak, $t_{\rm peak}\sim 3$ months, 
depends on $E$. For larger $E$, i.e., faster expansion, 
the rise starts earlier because of earlier
appearance of heating due to radioactivity; the decline after the peak is
earlier because of a larger velocity of the ejecta which causes the diffusion
time-scale to be shorter. The analytical treatment of this epoch is given in
detail by Imshennik \& Popov (1992). We caution that it is not just $E$ that
determines $t_{\rm peak}$ (for a given hydrogen distribution), but the
combination $E/M_{\rm env}$. $t_{\rm peak}$ is therefore mainly determined
by $E/M_{\rm env}$ and the hydrogen distribution.

Figure~\ref{bol3en} shows that both the first and the second parts of the 
light curve are reproduced well by 14E1 (see also Fig.~\ref{14elum3}).
14E1.3 is too bright near the minimum, while 14E0.7 is too dim near the
minimum and evolves too slowly.  Compared with the flux-limited diffusion model
(Shigeyama \& Nomoto 1990) 14E1 evolves similarly, but in much better
agreement with observations both near the minimum and near the peak when
a more accurate radiative transfer scheme is used.

For the progenitor we have used in Figure~\ref{bol3en} (i.e., the model with
mixing and $M_{\rm env} = 10.3$~\Msun), the best explosion energy in order to
fit the observations is close to $1.1 \times 10^{51}$~ergs. Considering the
uncertainties of the progenitor model in terms of mixing and envelope mass,
we obtain best fits to the bolometric light curve for the explosion energies
in the range $(0.85-1.35)\times 10^{51}$~ergs. As noted in Figure 9, 
the ``true'' bolometric is likely to be intermediate to the SAAO and CTIO/ESO
results used in our fits (see Suntzeff \& Bouchet 1990). This allows for 
a $\pm 10\%$ span in luminosity. However, a larger error in luminosity 
($\sim 15 \%$, Lundqvist et al. 1999b) is due to the still prevailing 
uncertainty in distance modulus to the supernova (e.g., Walker 1999), which 
gives a combined error of approximately $\pm 30\%$ in absolute luminosity.
The explosion energy should therefore be within the 
range $(0.8-1.4)\times 10^{51}$~ergs.

We note that $E$ is very similar to in the analytical diffusion models
of Arnett \& Fu (1989) and Imshennik \& Popov (1992). (The initial theory
of diffusion was developed by Arnett [1980, 1982] for Type Ia supernovae.)
All these models use an eigenvalue formulation of the problem, which is not
strictly correct (Blinnikov \& Popov 1993). However, the more correct, though
more complicated, moving-boundary formulation  produces
results which agree rather well with the results of Imshennik \& Popov
(1992; see Popov 1995). We therefore support the use of the results of
Imshennik \& Popov (1992) to make reliable estimates of the supernova
parameters from the near maximum light.

This rather limited range of energies we find is important for calculations 
of the ionization of the circumstellar gas (e.g., Lundqvist \& Fransson 1996;
Lundqvist 1999). Until now these calculations have been based on the models
500full1 and 500full2 of Ensman \& Burrows (1992). Qualitatively, the 500full1
model is rather similar to the models in our preferred energy range. We will
discuss this in detail in Lundqvist et al. (1999a), but from the analysis in
Lundqvist \& Fransson (1996) we note immediately that the ionization of the
outer rings is particularly sensitive to the spectrum of the burst.

\subsection{Radius of the Progenitor}

The above analysis has been based on a progenitor model with $R_0 =$
48.5 \Rsun.  The uncertainty in the  luminosity of the progenitor may imply
that $R_0$ could be uncertain by $\sim$ 20 \%  (\cite{woo88,snk88}).
Thus we have calculated the light curve models 14E1.26R, 14E1.34R, and 14E1.45R
for $R_0 =$ 40\Rsun, and one model, 14E1.4R6,  for $R_0 =$ 58\Rsun\
(Table \ref{runs} and Figs.~\ref{bol2r40}, \ref{bol2r4r6}).
The early light curve for the first 2 days is not so different from
the case of the standard $R_0=$ 48.5 \Rsun, but $t_{\rm prop}$ scales
as $R_0/E^{1/2}$.

However, the light curve near the minimum (plateau in $V$) is
dimmer for $R_0 =$ 40\Rsun\ because the luminosity at that epoch is
approximately proportional to $R_0$.
As a result, for $R_0 =$ 40\Rsun, we need an explosion energy larger
than $1.2 \times 10^{51}$ ergs to put the light curve in agreement with
the observations near the minimum (Fig.~\ref{bol2r40}). But then the peak
is too early. A larger radius, $R_0 =$ 58\Rsun, shows an opposite trend.
In this case,  one needs a lower energy for the brightness at the minimum, 
but then the peak is too late.

\subsection{Hydrogen mass}

We have also made some runs for models where the hydrogen abundance in the
outer layers was artificially raised to 0.7 (as is also assumed in, e.g.,
the non-evolutionary models by \cite{viu93}).
For example, the model 14E1.25H (Table \ref{runs}) has the same
composition of metals as 14E1, but H is enhanced at the expense of He.
So, the total mass of H is here $M_{\rm H}  \sim 7  \msun$,  while it
was $M_{\rm H}  \sim 5.5\msun$ in the standard runs (14E1, 14E1.3 etc.).
The hydrogen rich model 14E1.25H evolves similarly to
the standard runs 14E1.3 and 14E1.21 (Fig.~\ref{lb3heh}). This explains
the success of such models, as demonstrated by Utrobin (1993), but it is hard
to justify a solar H abundance for SN 1987A from the evolutionary point
of view, as well as in context of the abundances in the inner circumstellar 
ring (e.g., Lundqvist \& Fransson 1996).

\section{Conclusions}

In this paper we have described an extensive set of full radiation
hydrodynamics calculations aimed to improving the modeling of the first few
months of the light curve of SN 1987A. We have shown that the improved models
%rev
can reproduce the light curve and suggest that proper handling of the 
radiation transfer is indeed decisive for the success of model fits.
%rev

The full multi-group radiation hydrodynamic modeling is more reliable mainly
because the effects of scattering are treated self-consistently. Our findings 
are that:
\begin{enumerate}
\item the color temperatures and broad band photometry (and full continuum
spectra) are predicted correctly for the first days of the supernova.
\item the shape of the modeled light curve, especially near the bolometric
minimum at $\sim 10 $ days, and during the broad peak at $\sim 100$ days is
much better in the multi-group approach than in the equilibrium diffusion one.
\item the density profiles of the supernova at various epochs are smoother.
\end{enumerate}

Our code assumes LTE, but we have bracketed NLTE effects by extreme
assumptions of the treatment of spectral lines (as either being scattering
dominated, or fully absorptive). We find that the emission at shock break-out 
is not sensitive to those assumptions, which gives us confidence in our
results. This is supported by a very good agreement with {\it IUE} observations
for the first days. For the first 100 days, best agreement is obtained when 
NLTE effects are mimicked by treating the line opacity as absorptive,
following the prescription of Baron et al. (1996).

Looking at our results in greater detail, we find that the color temperature
around shock breakout exceeds $10^6$ K, which is higher than those obtained
using a more approximate approach (e.g., Shigeyama \& Nomoto 1990), but not so
different from the model 500full1 of Ensman \& Burrows (1992).
The large difference between color and effective temperatures persists for 
the first hours. This implies that the rise in the $V$ luminosity is slower 
than in equilibrium diffusion models (which was first noticed by H\"oflich 
1991). Because of the overwhelming dominance of electron scattering 
during the first hours, the rise in the $V$ band is too slow in all of our
models. This raises the question on the calibration of the first photometry
data of SN 1987A (McNaught \& Zoltowski 1987; West \& McNaught 1992).

It is indicated by our light curve modeling that mixing of \nifsx\ up
%rev
to $v\sim 3000 - 4000$~\kms~ could be needed, but our analysis is unlikely
to supersede those of spectral analysis (e.g., \cite{viuchuganna95},
Kozma \& Fransson 1998a,b), modeling of early X-ray emission (e.g., Kumagai
et al. 1989), or direct observations of infrared IR lines (Erickson et al.
1988).
%rev

We have improved on the constraining of $E$. The earlier flux-limited
diffusion calculations (Shigeyama \& Nomoto 1990) provided a constraint on $E$
from both the pre-peak light curve and the plateau-like maximum light,
concluding $E = (1.1 \pm 0.4) \times 10^{51}$ ergs. We find from our
more detailed analysis that the best agreement with the observations
is obtained for $E = (1.1 \pm 0.3) \times 10^{51}$ ergs. To arrive at this
result we have assumed that the most likely range of $M_{\rm env}$
is $M_{\rm env}$ = 7 - 10 $M_\odot$ (Saio \etal 1988b), in order for models of
the presupernova evolution to be consistent with the enhancement of N/C and N/O
in the circumstellar matter (Lundqvist \& Fransson 1996). Knowing the energy 
with this accuracy, as well as having a detailed spectroscopic evolution from 
our models, we can constrain the ionization of the circumstellar gas much 
better than before. This will be discussed in Lundqvist et al. (1999a).

\acknowledgments

We are grateful to Ron Eastman, Stan Woosley, Wolfgang Hillebrandt, Vladimir
Imshennik, Dmitriy Nadyozhin, Nikolai Chugai, Cecilia Kozma,
Alexandra Kozyreva, Jason Pun and Victor Utrobin for discussions, and also
Ron Eastman, Vlad Popolitov and Elena Sorokina for using their computer
codes as part of ours.

Preliminary results of this work were reported at the ``CTIO/ESO/LCO Workshop
on SN 1987A: Ten years after'' in the contributions by Lundqvist \& Sonneborn
(1999) and Nomoto, Blinnikov, \& Iwamoto (1999). We would like to thank the
participants of this workshop for stimulating discussions.

This work was supported in Russia by a grant from The International Science \&
Technology Center 97-370, and by the Russian Basic Research Foundation grants
RBRF 96-02-19756 and RBRF 96-02-17604. We are also grateful to grants from
The Royal Swedish Academy of Sciences and The Wenner-Gren Center Foundation for
Scientific Research. P.L. receives further support from The Swedish Natural
Science Research Council and The Swedish Board of Space Research. The project
has in part also been supported by the grant-in-Aid for Scientific Research 
(05242102, 06233101) and COE research (07CE2002) of the Ministry of the 
Education, Science, Culture, and Sports in Japan.

%\end{references}

\clearpage

\figcaption[]{Density as a function of the interior mass, $M_{\rm r}$, and
of the radius $r$ in our remap of the presupernova model 14E1
(compare with Fig.~1 in %with  mass 16.27 \Msun
\protect\cite{shi90}). Mass cut is at $M_{\rm c}=1.6 \msun$.
\label{nommxfrho}}

\figcaption[]{Composition as a function of interior mass,
$M_{\rm r}$, for the most abundant elements in the unmixed presupernova model
from Nomoto \& Hashimoto (1988) used for runs 14E1U and 14E1.2U.
Mass cut is at $M_{\rm c}=1.6 \msun$.
\label{unmxchm}}

\figcaption[]{Abundance distribution as a function of enclosed mass
for the ejecta in model 14E1 with mixing.
\label{nommxfchm}}

\figcaption[]{The time for the shock to reach the surface of the star according
to Eqn.\protect\ref{tprop} versus the computed time for the models in
Table \protect\ref{earlight}. Dashed line shows direct proportionality between
the two times.
\label{tmre}}

\figcaption[]{Density profile against radius at $t=101$ days for model
14E1.
\label{rholgr}}

\figcaption[]{Changes in the velocity profile near shock breakout for model
14E1.3. The figure shows how the outermost ejecta are accelerated to 
homologous expansion.
\label{nomv}}

\figcaption[]{Very early bolometric light curve, and color and effective
temperatures for the run 14E1. Realistic, scattering dominated opacity
has been assumed. Solid line shows the temperature of the best blackbody fit 
to the flux ({\sl color} temperature). Dashed line shows the effective
temperature defined by the luminosity and the radius of last scattering.
\label{nommxftphea}}

\figcaption[]{Spectral flux in observer's frame at first maximum light
for models 14E1 (i.e., with realistic scattering; dashed line) and 
14E1A (i.e., with forced absorption; solid line).
\label{max1st}}

\figcaption[]{Early bolometric luminosity for runs 14E1.3 (dashed), 14E1 
(solid), 14E0.7 (short-dashed). Observations are from data obtained at SAAO
(squares, Catchpole \etal 1987) and CTIO (crosses Hamuy \etal 1988).
%rev
Note that we in figures showing the bolometric light curve have chosen to 
show both the SAAO and CTIO/ESO results. The difference between the two
photometric systems results in a maximum difference in bolometric luminosity 
of $\sim 1.2$ As argued in Suntzeff \& Bouchet (1990), the ``true''
result is likely to be intermediate to the SAAO and CTIO/ESO results.
(See also \S 7.1.)
%rev
\label{14elum3}}

\figcaption[]{Absorption ($\alpha$, solid) and scattering ($\sigma$, dotted) 
opacity as a function of
wavelength for three different temperatures. Density is
the same in all plots, $\rho=10^{-10}$ \gmc. This density is typical for
the photospheric layer at the peak luminosity at shock breakout.
Note the dominance of scattering for $T\ga 3\times 10^4$ K.
%rev
\label{scaabs}}

\figcaption[]{Same as Figure~\protect\ref{14elum3} but for the color
temperature $T_{\rm c}$. $T_{\rm c}$ was derived from a best-fitting blackbody
to the modeled spectrum.
\label{14etph3}}

\figcaption[]{Apparent $V$ magnitude in observer's frame for model 14E1.21. A
reddening by $E_{B-V}=0.15$ (Wampler, 1988) was applied, and the distance
modulus was assumed to be 18.5. Jones' limit is shown by
inverted {\sf Y}. ``1987'' marks m = $6.36 \pm 0.11$ measured by
McNaught \& Zoltowski (1987), and ``1992'' is m = $5.94 \pm 0.10$ found by
West \& McNaught (1992). The filled circles are the NLTE predictions
by H\"oflich (1990, 1991), who took the temperature structure from the 14E1.25
model computed in the flux-limited equilibrium diffusion approximation
by Shigeyama \& Nomoto (1990).
\label{vearly}}

\figcaption[]{$V$ light curve for the first day for two values of the initial
radius $R_0$ = 48.5 \Rsun\ and 40 \Rsun\ (i.e., models 14E1.3 and 14E1.34R,
respectively.)  The `old' reductions of the observations are given according
to the compilation in Arnett (1988) and the `new' reductions are from 
West \& McNaught (1992) and Shelton \& Lapasset (1993).
\label{eashv2}}

\figcaption[]{Early $UBV$ fluxes in observer's frame for 14E1. A reddening
by $E_{B-V}=0.15$ and a distance modulus of 18.5 have been applied.
The uppermost curve is the $V$ flux for 14E1A (forced absorption). Jones'
limit is shown by inverted {\sf Y}.
\label{nommxfea}}

\figcaption[]{Change in the temperature profile against $M_{\rm r}$ for 14E1.
Solid and dotted lines show gas and radiation temperatures,
respectively.  The attached numbers show the time in days after the
explosion. Note that the time here is the comoving time, and not the 
retarded, as in graphs for observations. The hydrogen recombination front is
rather broad and propagates inward in $M_{\rm r}$.
\label{settm}}

\figcaption[]{Same as Figure~\protect\ref{settm} but only for the material
temperature against radius $r$.
\label{settr}}

\figcaption[]{Same as Figure~\protect\ref{settm} but for the comoving luminosity
$L$. The drop in $L$ in the outer layers is explained by the fast motion of 
those layers (compare to the $L(r)$ graph in Fig.~17).
Another reason for the drop in $L$ is the effect of retardation. For $t=105$
days the outer layers `remember' the higher luminosity some time ago.
\label{setlm}}

\figcaption[]{Comoving luminosity $L$ against radius $r$ in linear scale for 
days 59 and 105.
\label{setlr}}

\figcaption[]{Bolometric light curves for 14E1 (with mixed composition), 14E1M
(mild mixing of \nifsx) and the unmixed model 14E1U. Squares are the data of
Catchpole \etal (1987), crosses - Hamuy \etal (1988).
\label{nommxfbol}}

\figcaption[]{Abundance distribution as a function of expansion velocity for
the model 14E1 with mixing.
\label{14e1vchm}}

\figcaption[]{Distribution of H and \nifsx\ as a function of expansion velocity
in the unmixed model 14E1U (solid and short-dashed lines) and in 14E1M
(mild mixing of \nifsx, dashed and dotted lines).
\label{14e1uvch}}

\figcaption[]{$UBV$ magnitudes in observer's frame for model 14E1. A reddening
of $E_{B-V}=0.15$ and a distance modulus of 18.5 have been applied.
\label{14e1ubv}}

\figcaption[]{Same as in Figure~\protect\ref{14e1ubv} but for the run 14E1S in
which the line opacity is scattering dominated.
\label{14e1subv}}

\figcaption[]{Predicted fluxes in {\it IUE} bands for model 14E1 during the 
first 10 days (solid lines). The fluxes are in observer's frame, and the 
distance modulus has been set to 18.5. The observational data of  Pun et al. 
(1995) are dereddened.
\label{iue10}}

\figcaption[]{Same as in Figure~\protect\ref{iue10} but for the first 100 days.
\label{iue100}}

\figcaption[]{Best blackbody fit temperature (solid line), and effective 
temperature (dotted) for 14E1. Squares are data of Catchpole \etal (1987), 
crosses -- Hamuy \etal (1988).
\label{tpfit}}

\figcaption[]{Radius of the ``photosphere'', $R_{\rm obs}$, found from the 
blackbody fit temperature for 14E1 (solid line). Dashed line shows the radius
$R_{2/3}$, where the optical depth in the continuum is $\approx 2/3$
at $\lambda\approx 5000$ \AA. Squares are data of Catchpole \etal
(1987), crosses - Hamuy \etal (1988) for $R_{\rm obs}$.
\label{rfit}}

\figcaption[]{Matter velocity at the ``photosphere'', $R_{\rm obs}/t$, and at
the optical depth $2/3$, $R_{2/3}/t$ (dashed), for 14E1. Squares are data of
Catchpole \etal (1987), crosses - Hamuy \etal (1988). Circles are the
values of $v$ found from Fe~II~$\lambda$5169 by \cite{phil88}.
\label{vphot}}

\figcaption[]{Bolometric light curves for different explosion energies: 14E1.3
(dashed), 14E1 (solid), 14E0.7 (short-dashed) for the mixed composition in
Figure~\protect\ref{nommxfchm}. Squares are data of Catchpole \etal (1987),
crosses - Hamuy \etal (1988).
\label{bol3en}}

\figcaption[]{Bolometric light curves for $R=40R_\odot$ for runs 14E1.26R and
14E1.45R, which both have mixed composition.
\label{bol2r40}}

\figcaption[]{Bolometric light curves for $R=40R_\odot$ for run 14E1.45R,
and $R=58R_\odot$ for run 14E1.4R6.
\label{bol2r4r6}}

\figcaption[]{Bolometric magnitude for model 14E1A (forced absorption instead
of scattering, solid) and 14E1S (scattering lines, dashed).
Squares are data of Catchpole et al. (1987), crosses - Hamuy et al. (1988)
\label{nommxfabbol}}

\figcaption[]{Bolometric light curves for the hydrogen~rich model 14E1.25H,
and the standard models 14E1 and 14E1.21.
\label{lb3heh}}

\vfill

%%%%%%%%%%%%%%%%%%%%%%%%%%% Tables %%%%%%%%%%%%%%%%%%%%%%%%%%%%%%%%%%%%%%%
\begin{table}
\caption{Runs\label{runs}}
\begin{center}
\begin{tabular}{lllllll}
\tableline
\tableline
run & $M_{\rm ej}/\msun$ & $R_0/R_\odot$ & $E_{\rm 51}$ & 
forced $\chi_{\rm abs}$ & Mixed H &   \\
\tableline
14E0.7  & 14.67 & 48.5 &  0.72   & no  & yes    \\ % nommxfhe14
14E1    & 14.67 & 48.5 &  1.03   & no  & yes    \\ % nommxfhe17
14E1.21 & 14.67 & 48.5 &  1.21   & no  & yes    \\ % 14E1_25
14E1.3  & 14.67 & 48.5 &  1.34   & no  & yes    \\ % nommxfhe
14E1M   & 14.67 & 48.5 &  1.04   & no  & yes     \\ %
14E1U   & 14.67 & 48.5 &  1.01   & no  & no     \\ % unmx15
14E1.2U & 14.67 & 48.5 &  1.20   & no  & no     \\ % unmx
14E1S   & 14.67 & 48.5 &  1.03   & no  & yes    \\ % 14e1s
14E1A   & 14.67 & 48.5 &  1.01   & yes  & yes   \\ %  14E1A 14E1Atau
14E1.25H & 14.67 & 48.5 &  1.25   & no  & yes    \\ % nm3mxfabse H rich
14E1.26R & 14.67 & 40  &  1.26    & no  & yes    \\ % nmr40he
14E1.34R & 14.67 & 40  &  1.34    & no  & yes    \\ % nmr40he208
14E1.45R & 14.67 & 40  &  1.45    & no  & yes    \\ % nmr40he22
14E1.4R6 & 14.67 & 58  &  1.42    & no  & yes    \\ % nmr60he2 R=58
\tableline
\end{tabular}
\end{center}
\end{table}

%\clearpage
\begin{table}
\caption{Predictions for the first maximum light \label{earlight}}
\begin{center}
\begin{tabular}{llllllll}
\tableline
\tableline
run & $t$, day & $L_{\rm bol}$, ergs s$^{-1}$  & $T_{\rm c}$, K
     & $T_{\rm eff}$, K & $R_{\tau=2/3}$, cm & $\int_0^{2 {\rm =
d}}Ldt$, ergs\\
\tableline
14E0.7     & .08960 & 4.217\e{44} & 1.074\e{6} & 4.71\e{5} &
3.32\e{12}&1.07\e{47}\\
14E1       & .07637 & 6.751\e{44} & 1.219\e{6} & 5.28\e{5} &
3.32\e{12}&1.40\e{47}\\
14E1.3     & .06726 & 9.466\e{44} & 1.339\e{6} & 5.73\e{5} &
3.32\e{12}&1.77\e{47}\\
14E1U      & .07692 & 6.616\e{44} & 1.207\e{6} & 5.24\e{5} &
3.32\e{12}&1.41\e{47}\\
14E1.2U    & .07161 & 7.939\e{44} & 1.268\e{6} & 5.49\e{5} &
3.32\e{12}&1.58\e{47}\\
14E1.26R   & .05768 & 8.665\e{44} & 1.428\e{6} & 6.18\e{5} &
2.74\e{12}&1.30\e{47}\\
14E1.34R   & .05620 & 9.175\e{44} & 1.451\e{6} & 6.26\e{5} &
2.74\e{12}&1.36\e{47}\\
14E1.45R   & .05389 & 1.012\e{45} & 1.497\e{6} & 6.42\e{5} &
2.74\e{12}&1.44\e{47}\\
14E1.4R6   & .07520 & 1.013\e{45} & 1.267\e{6} & 5.30\e{5} &
3.95\e{12}&2.38\e{47}\\
\tableline
\end{tabular}
\end{center}
Note. Maximum $T_{\rm eff}$ almost coincides with the peak
$L_{\rm bol}$, while maximum $T_{\rm c}$ is $\sim 100$ s earlier.
\end{table}

\vfill 

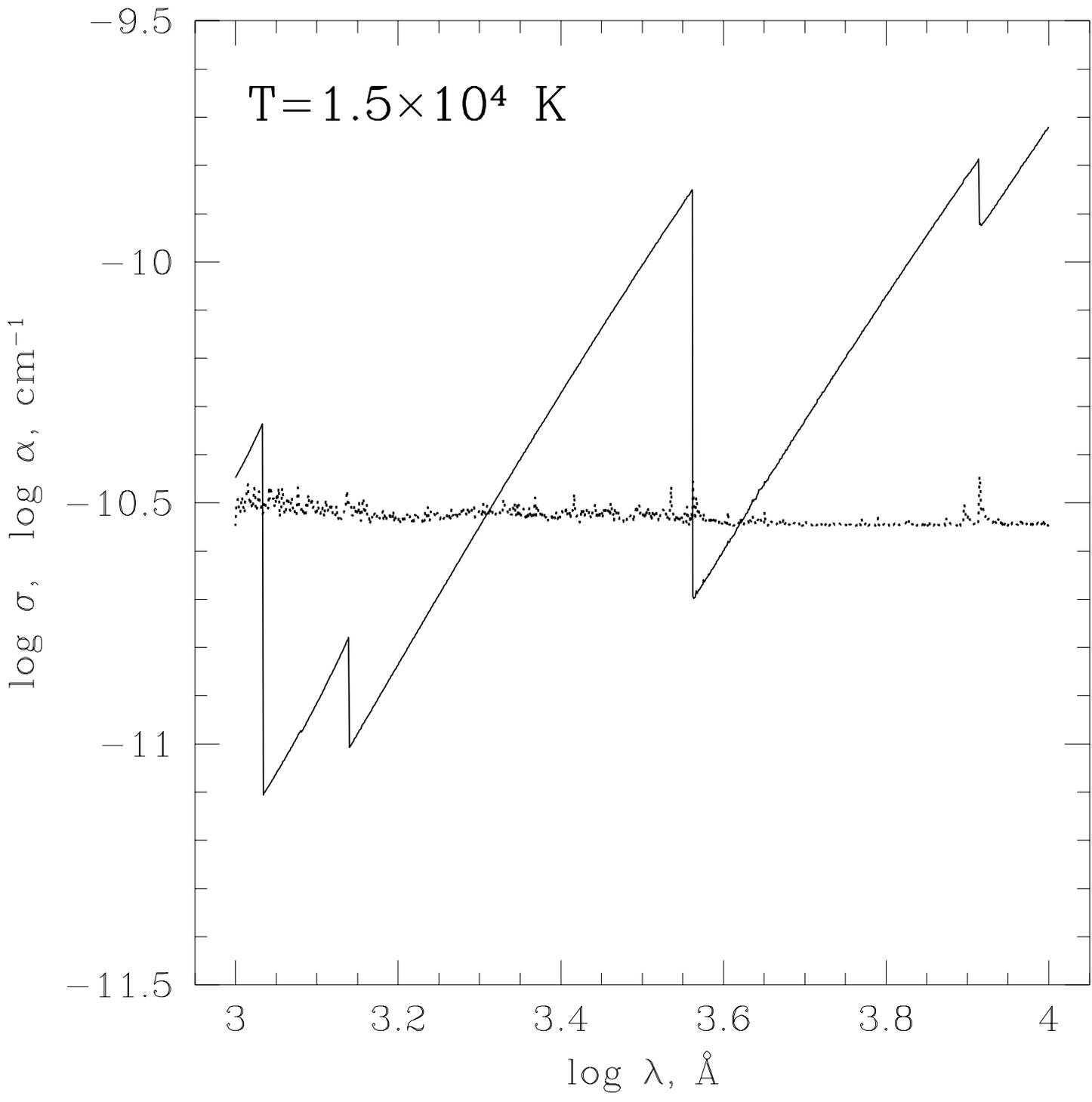
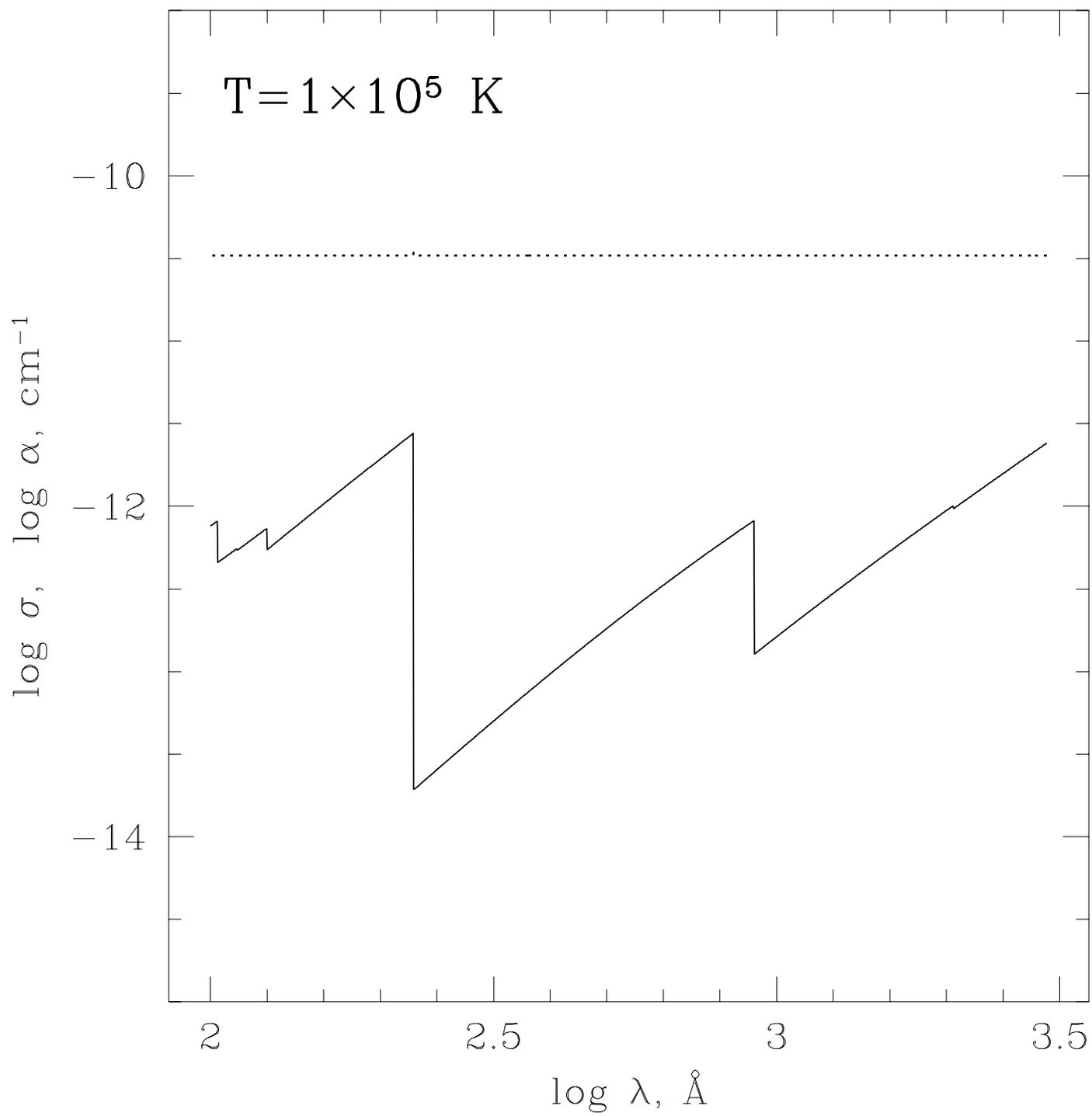
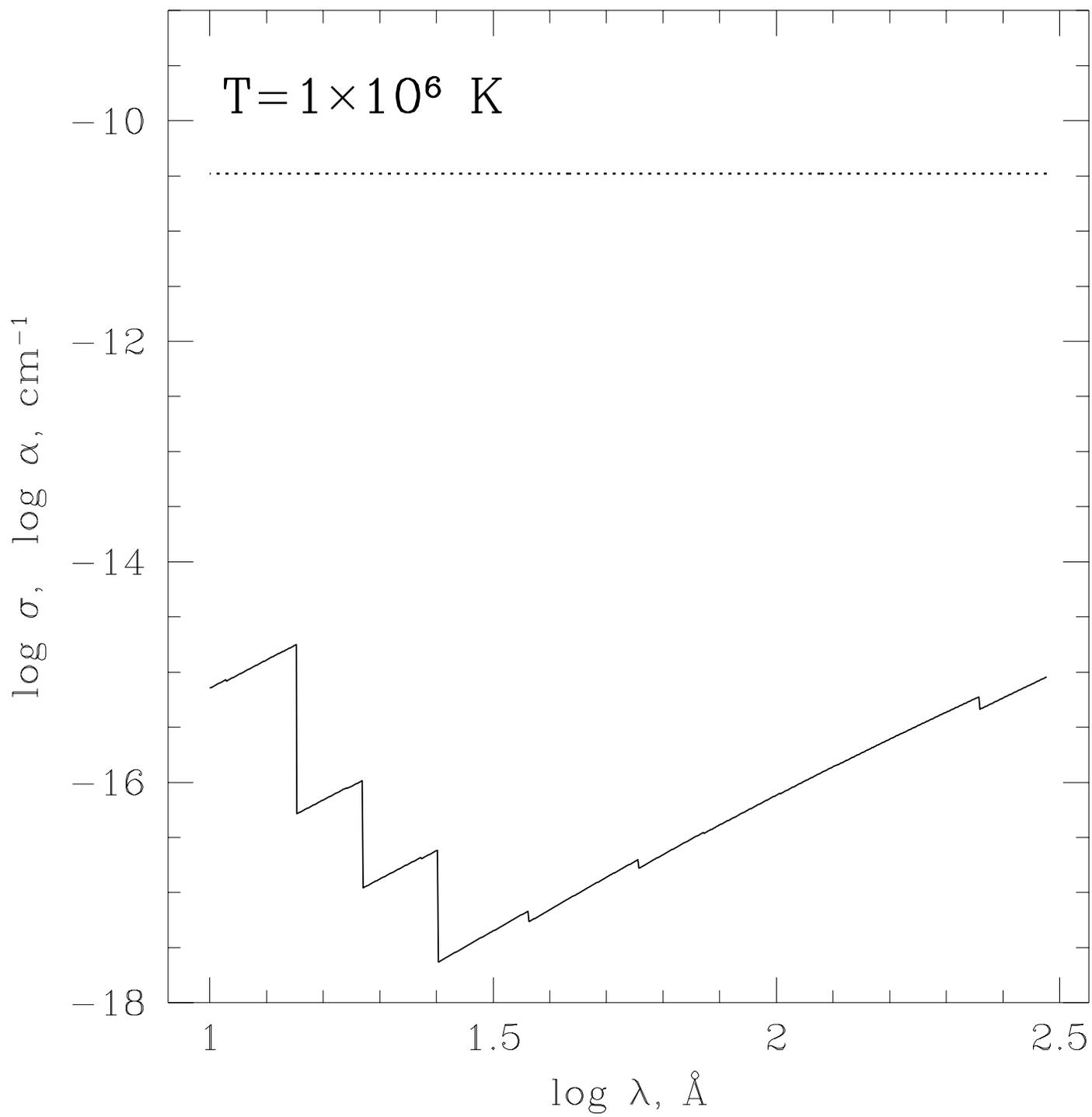
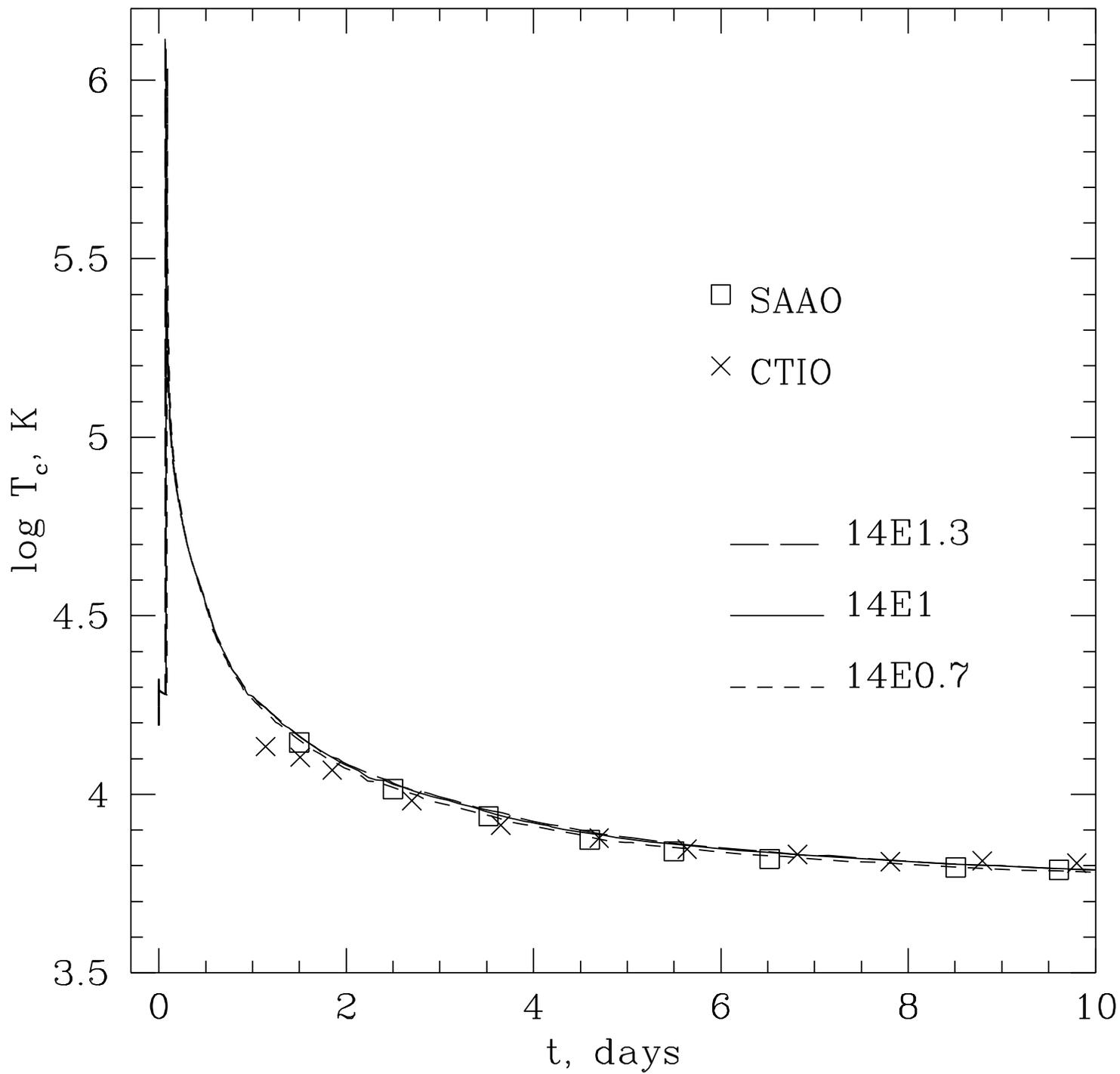
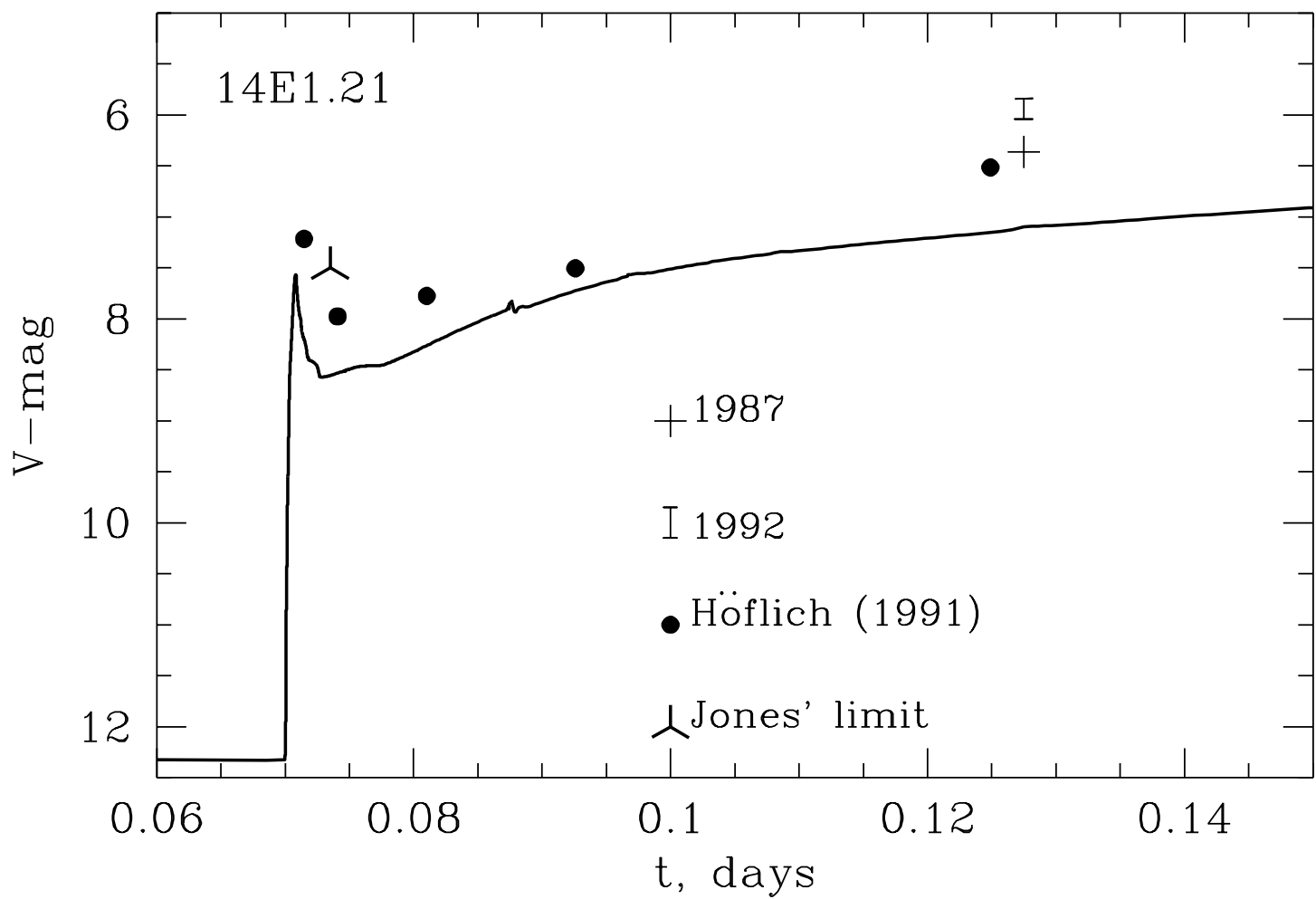
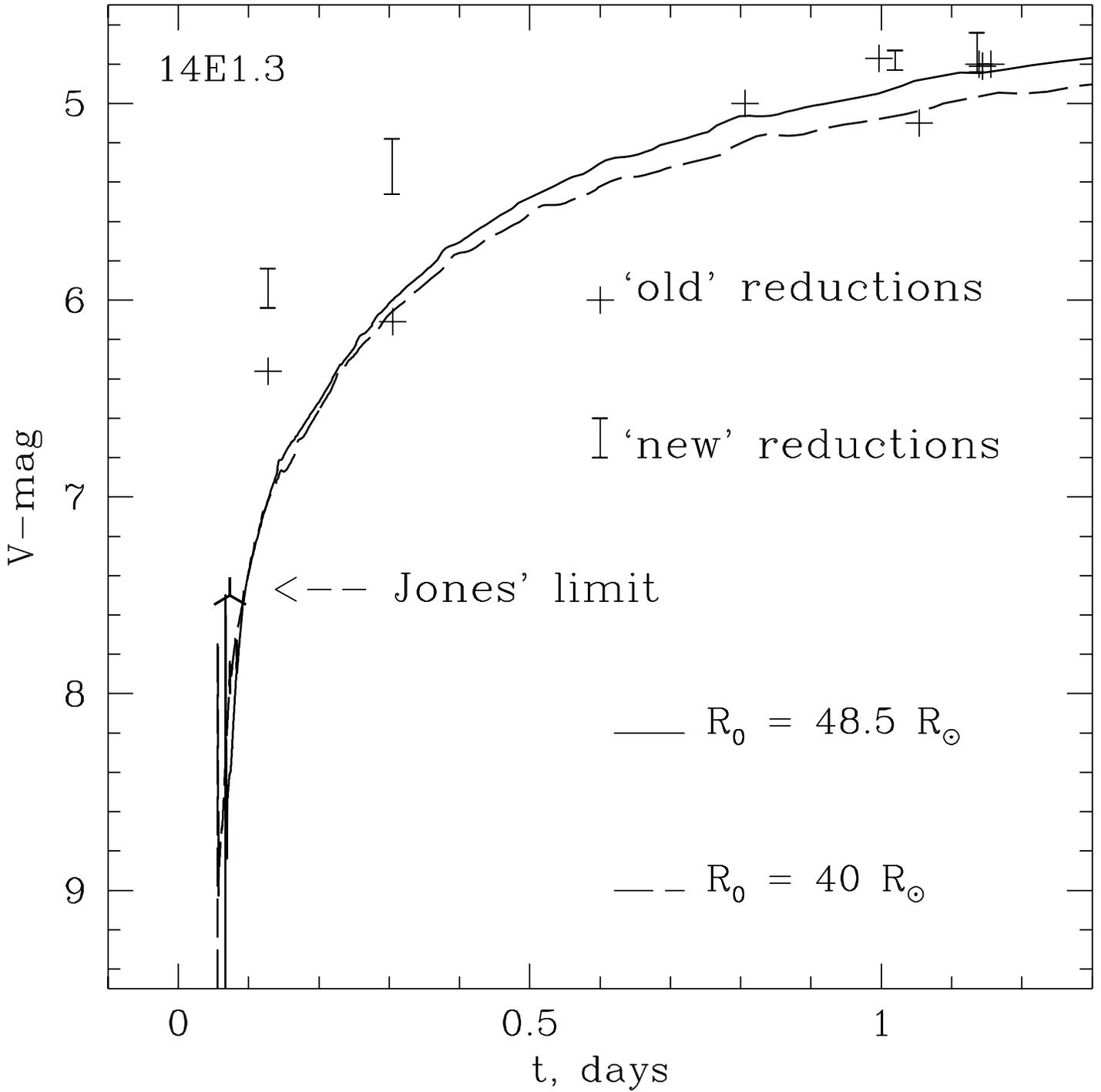
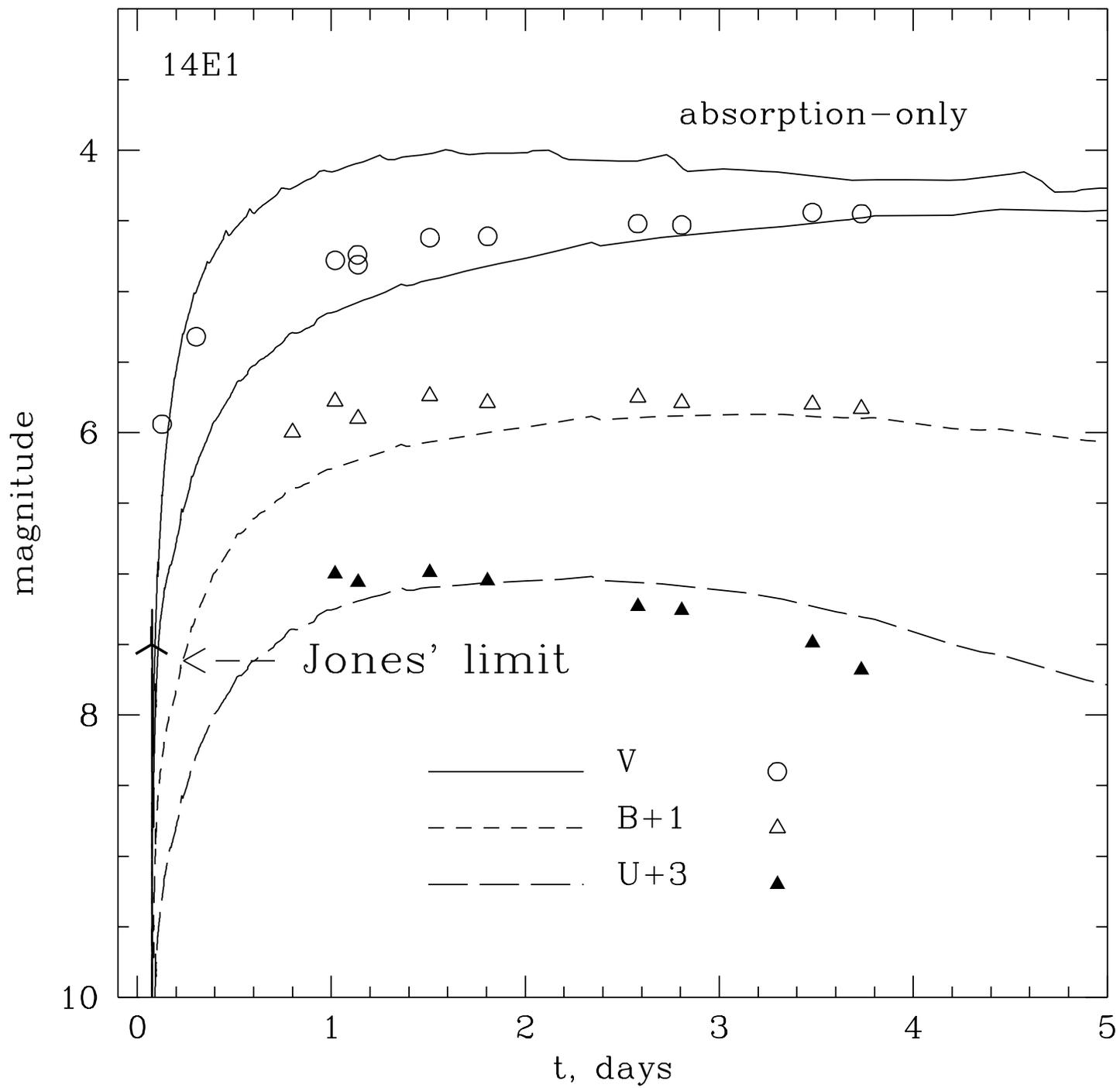
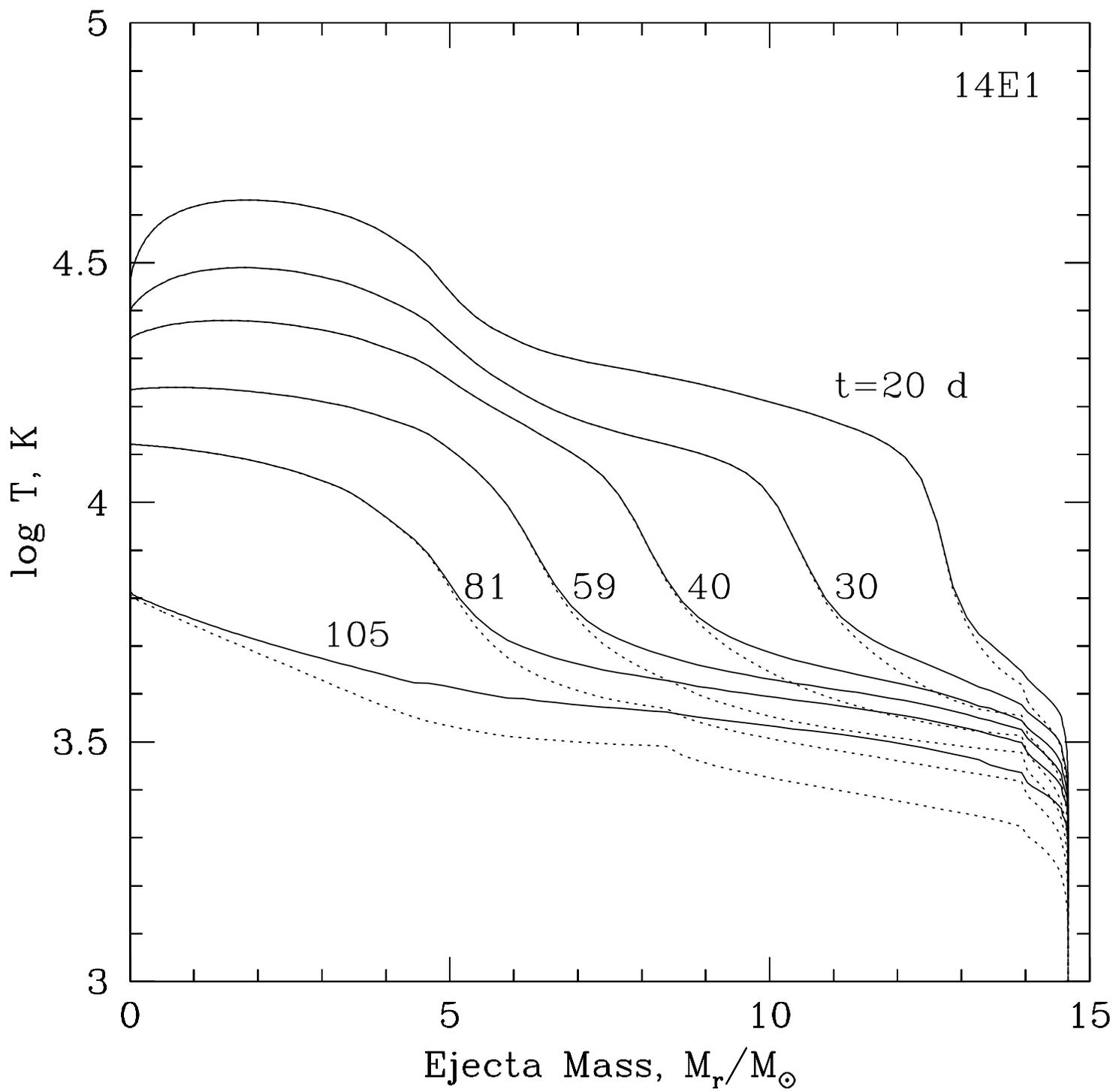
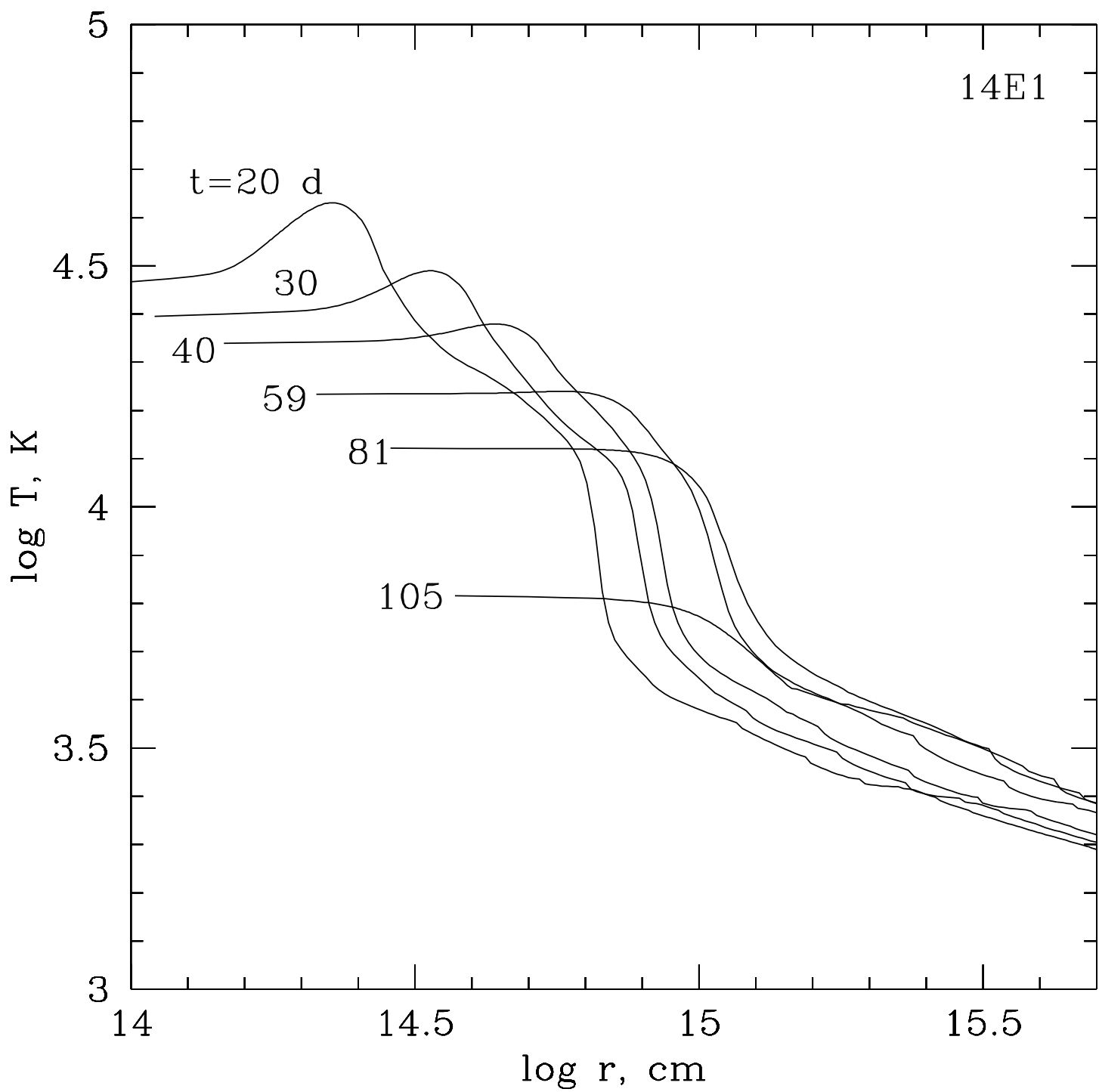
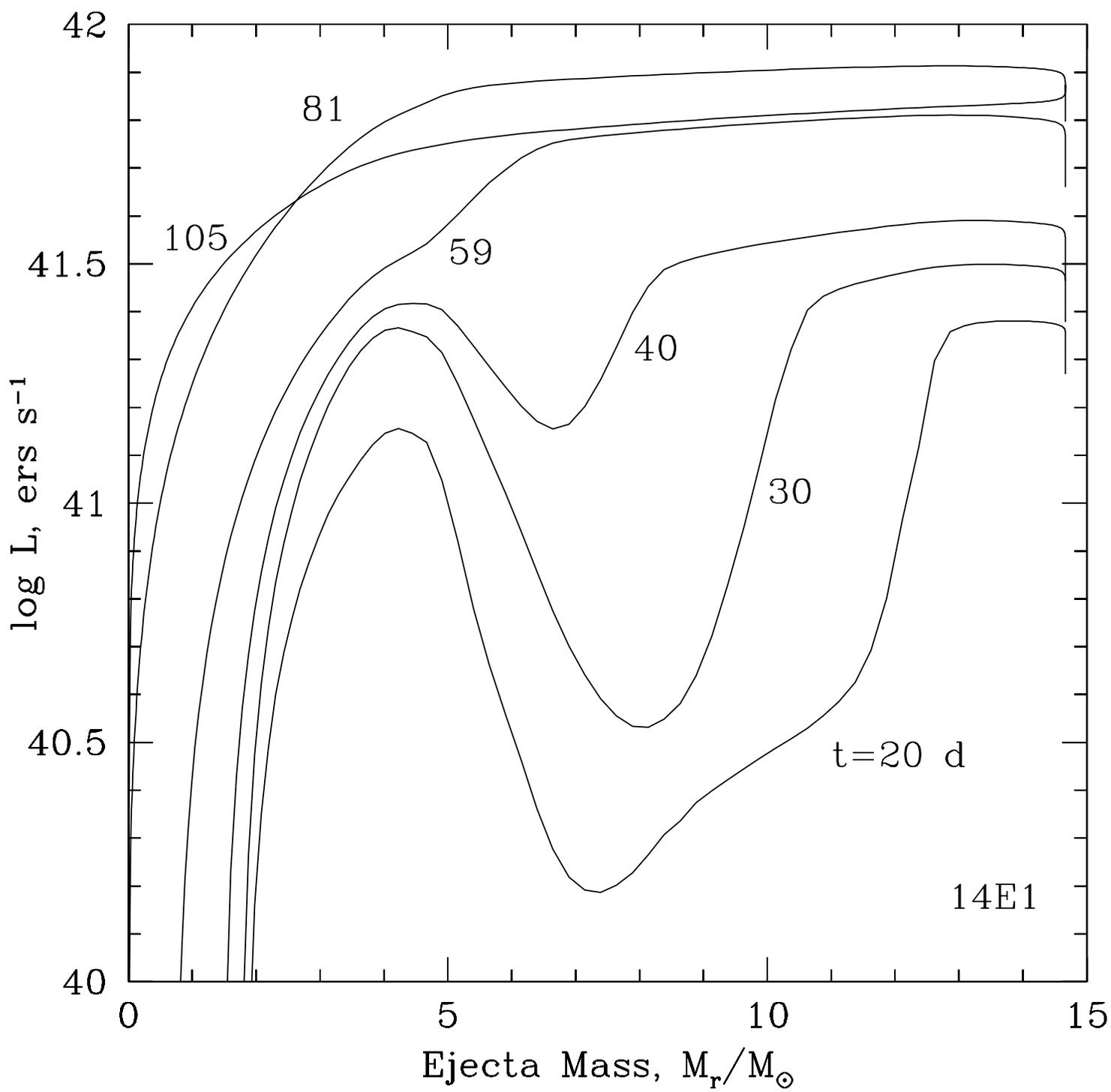
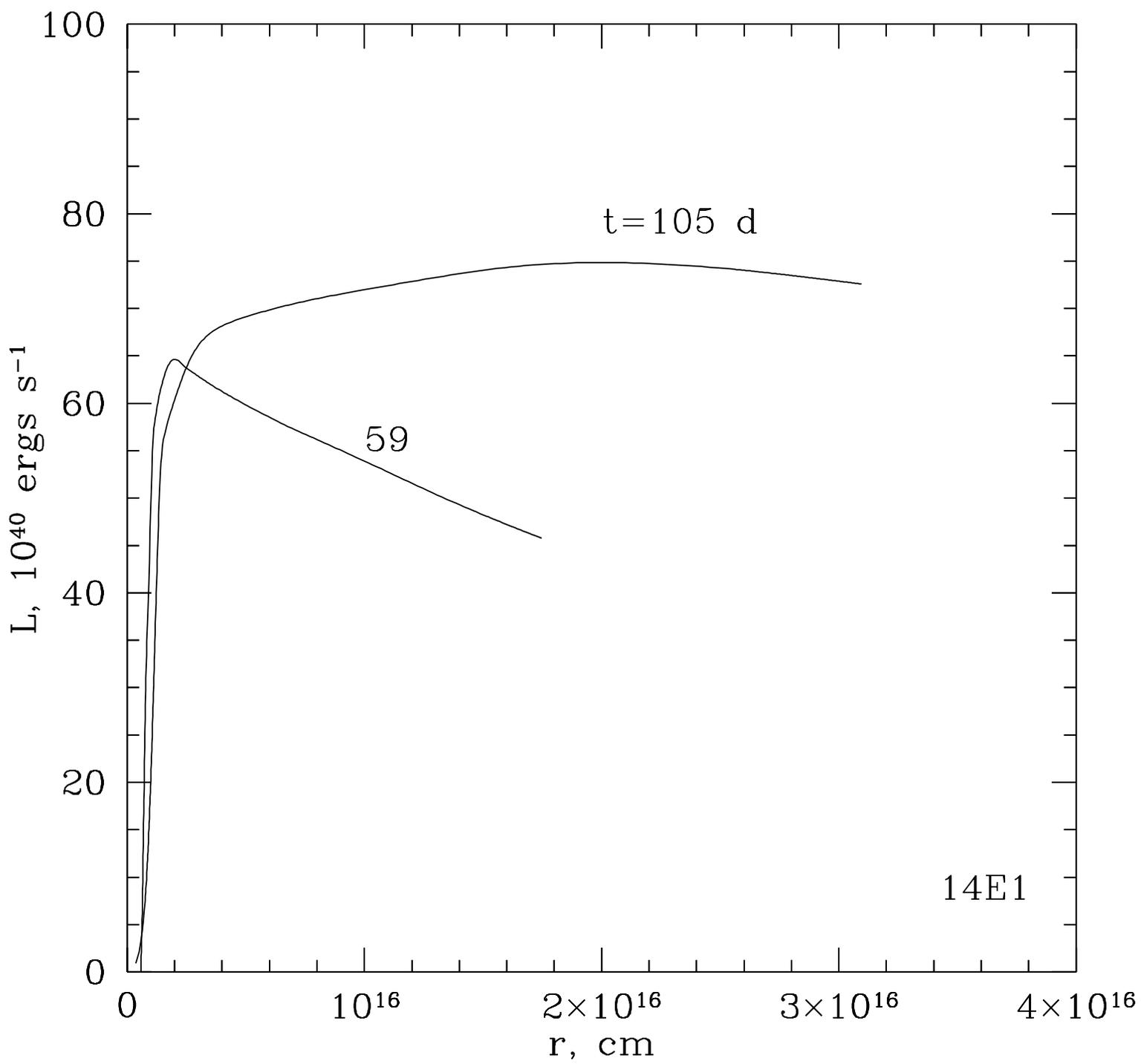
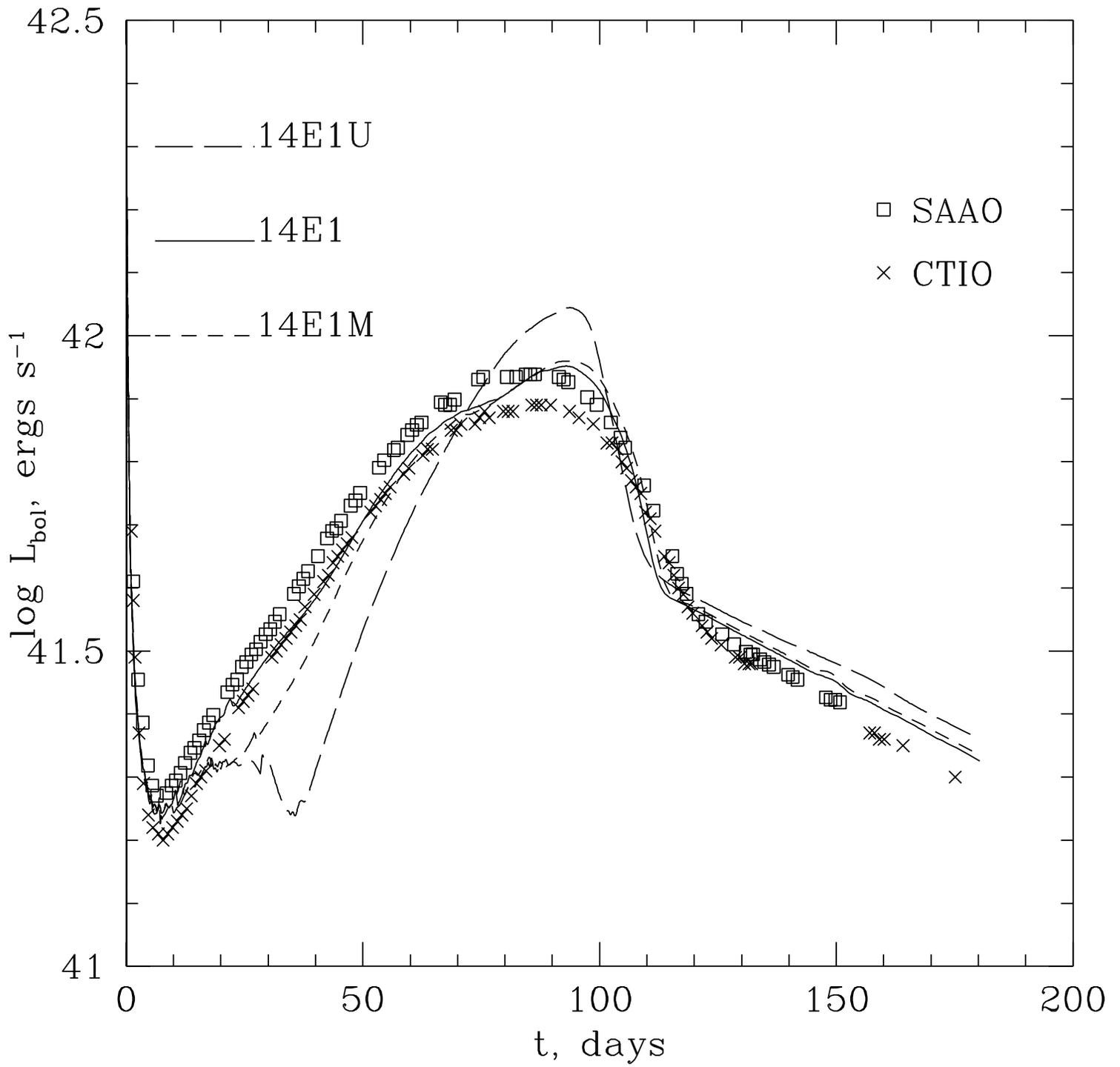
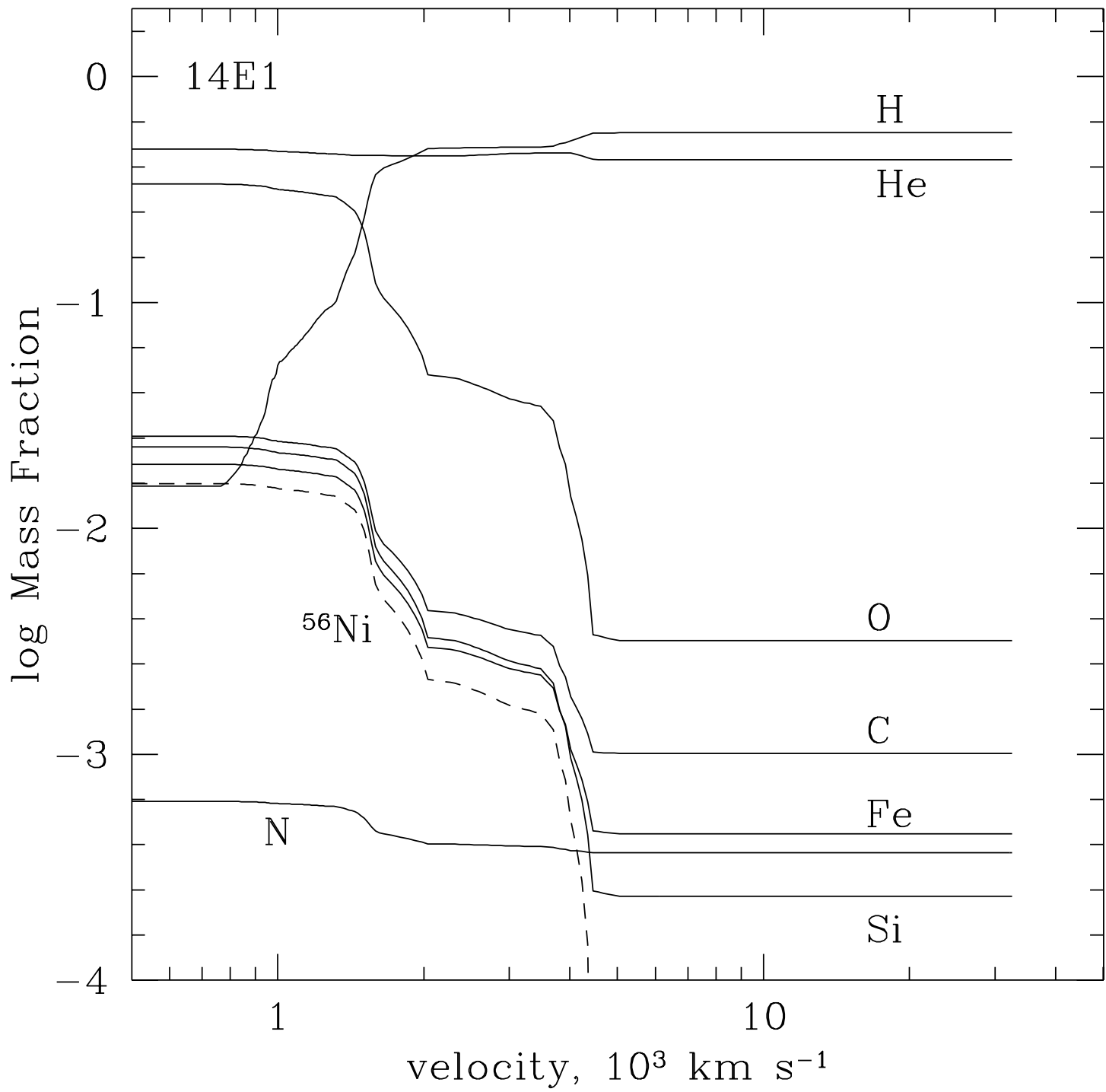
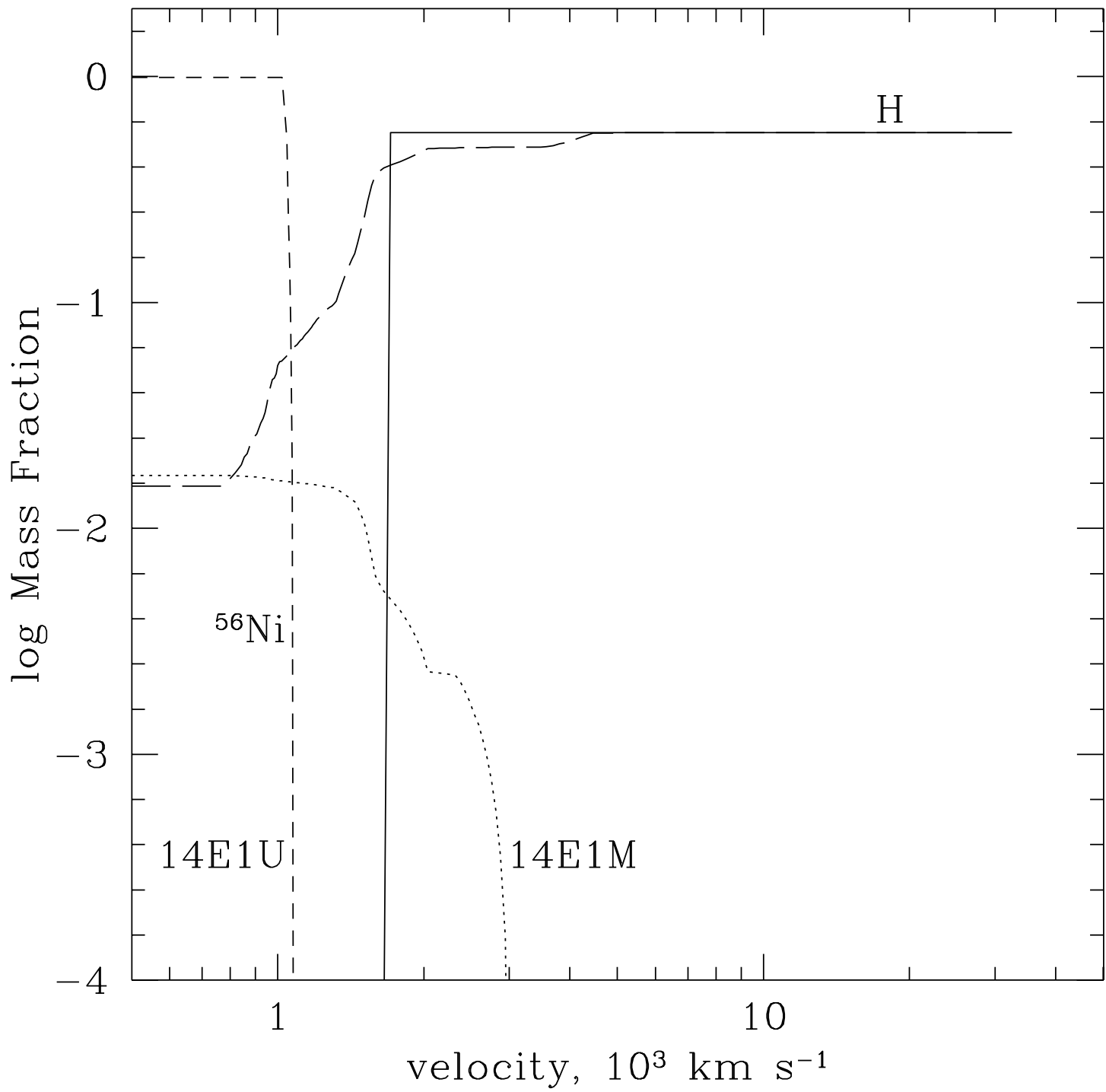
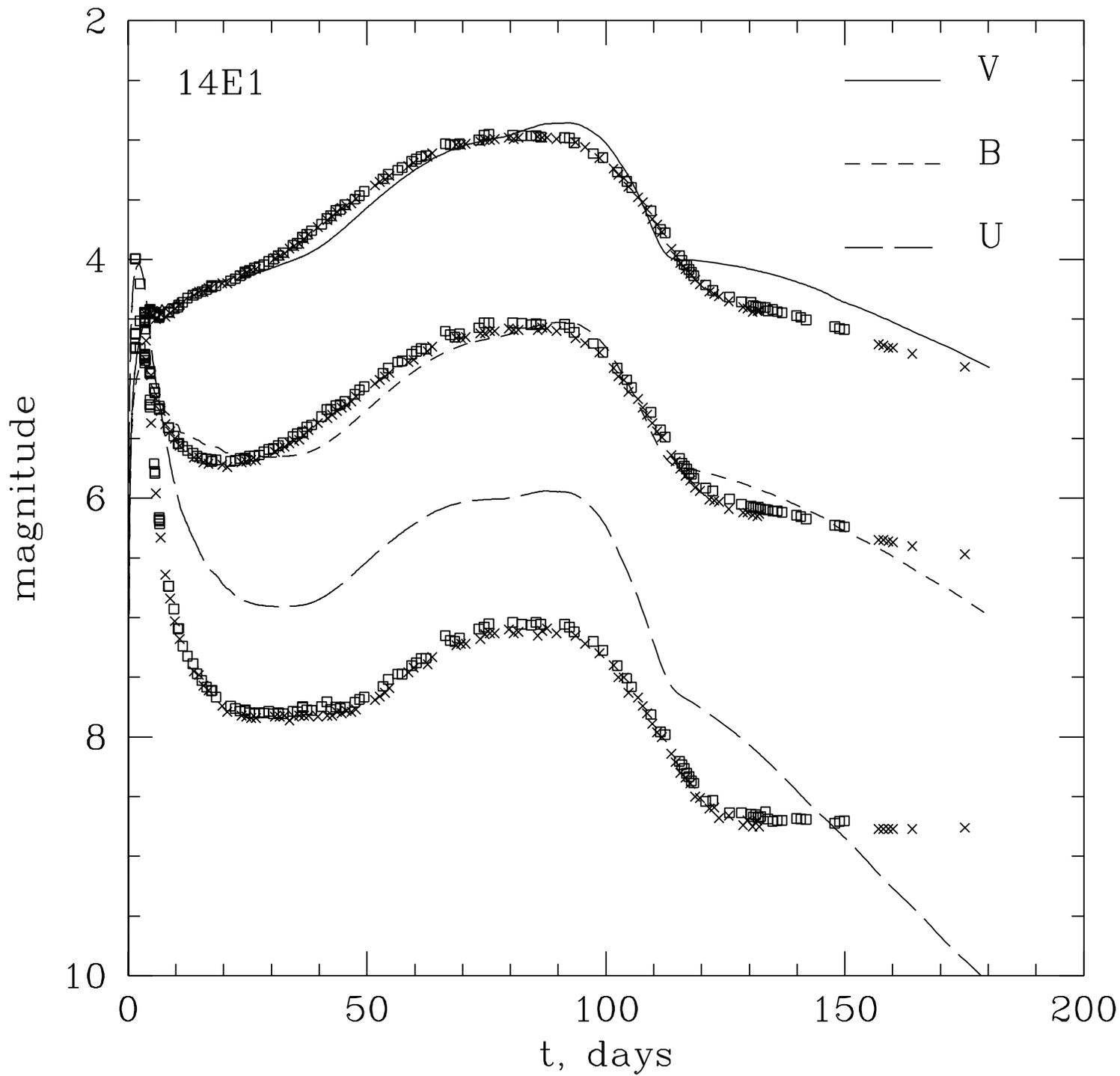
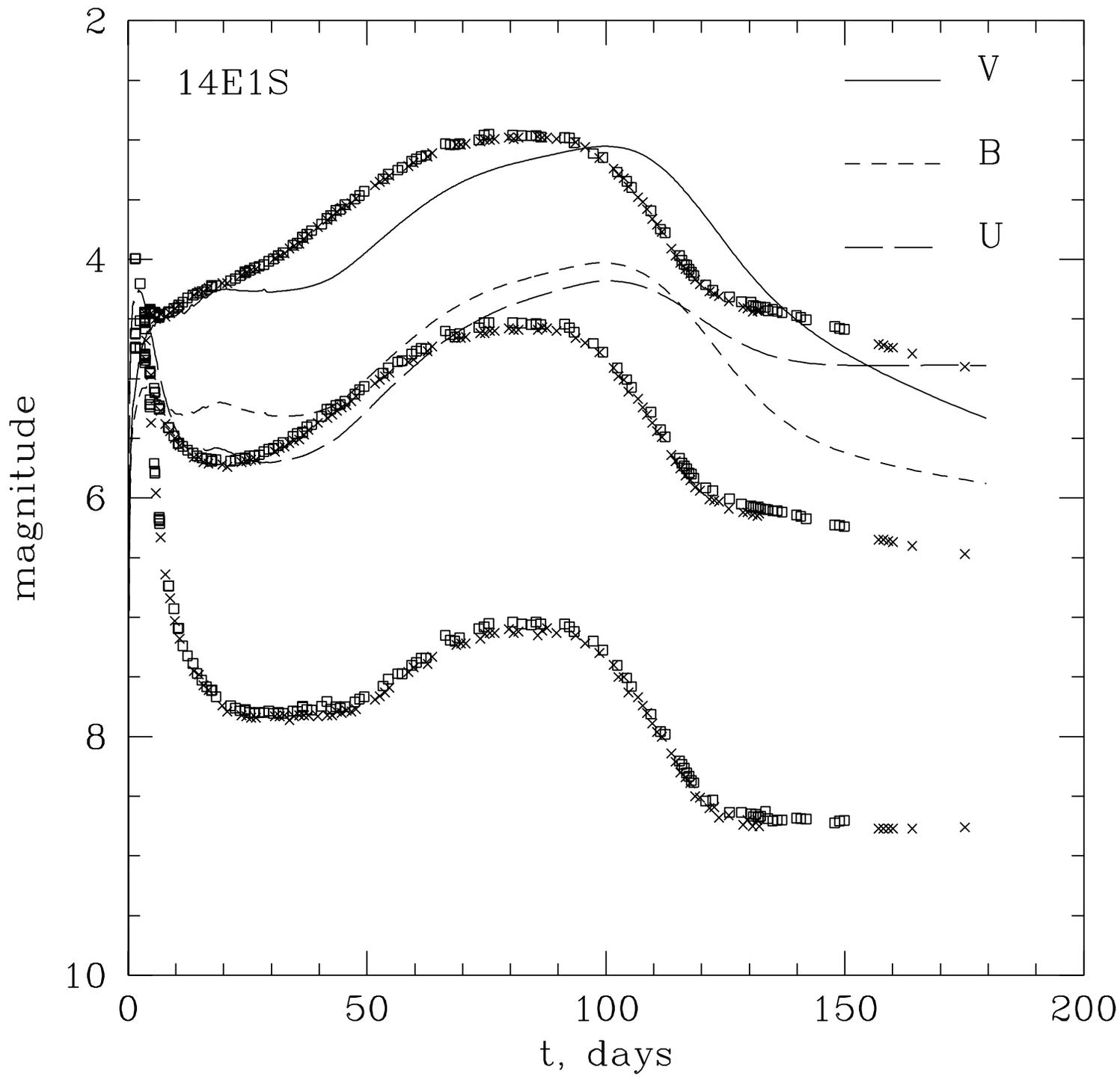
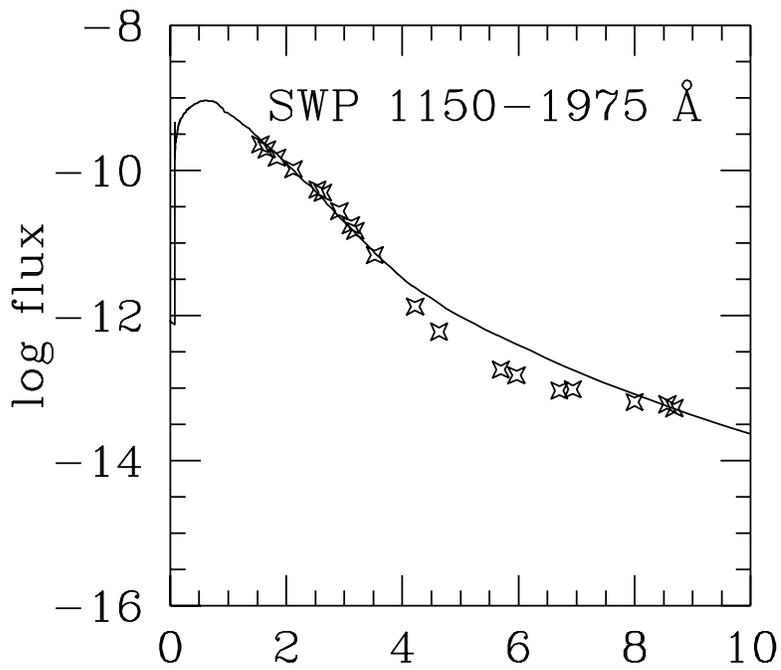
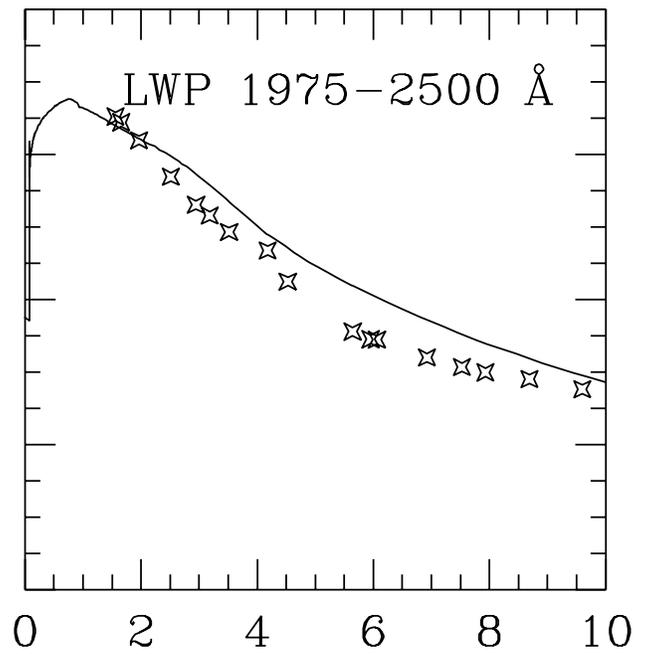
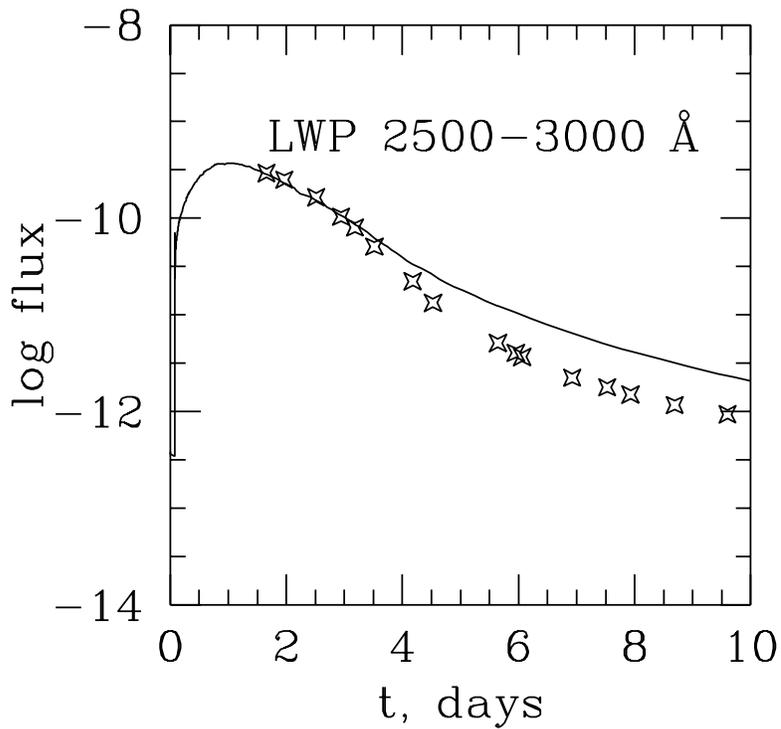
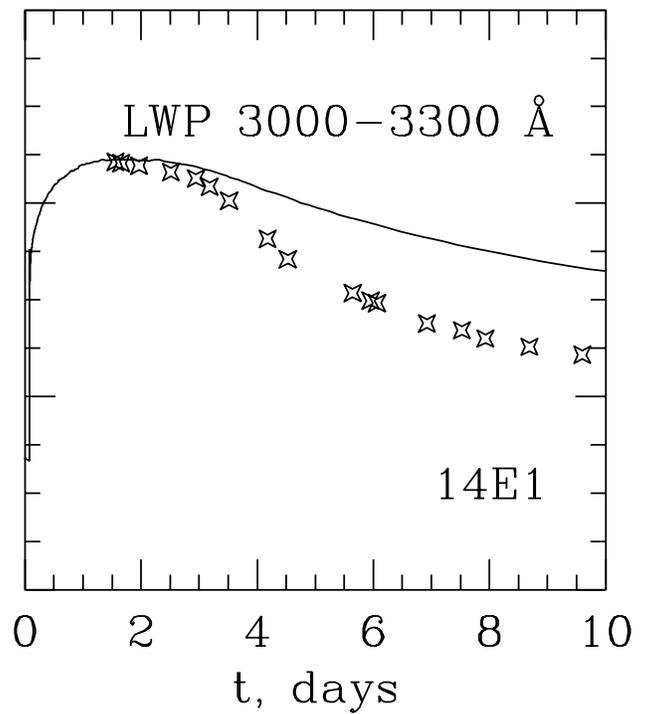
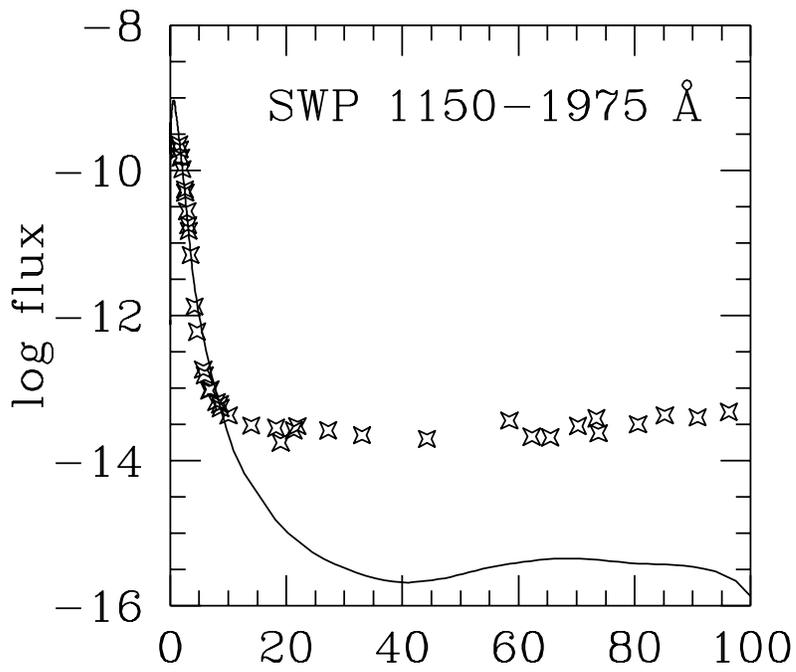
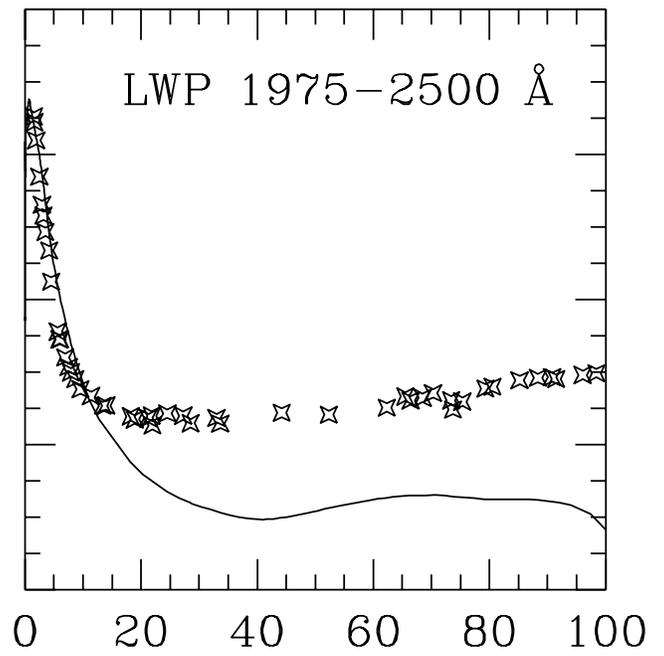
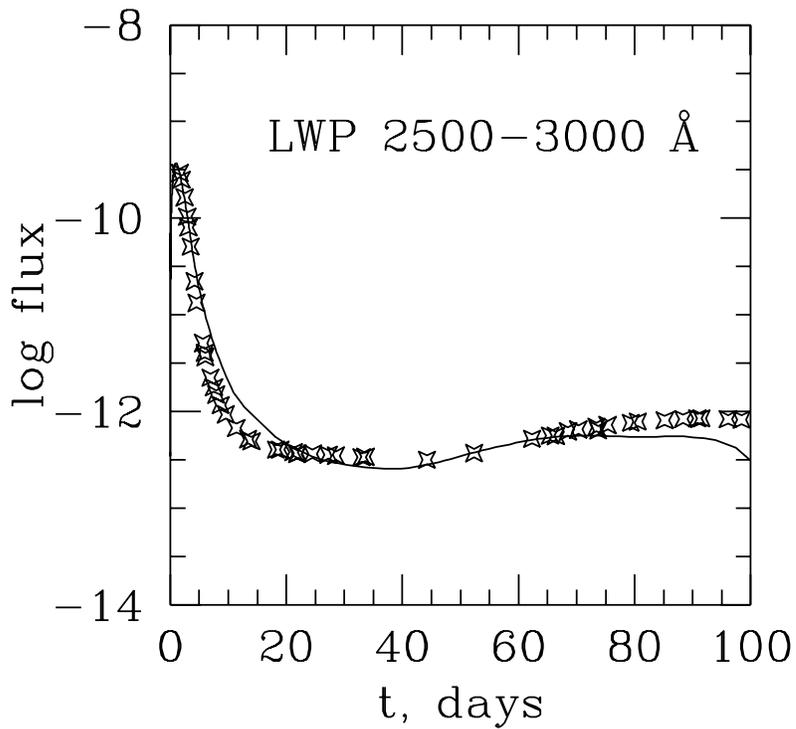
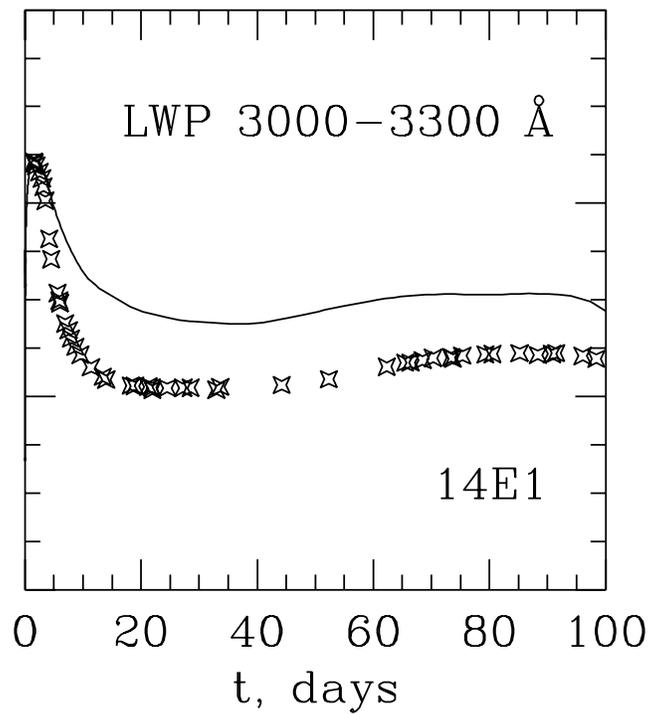
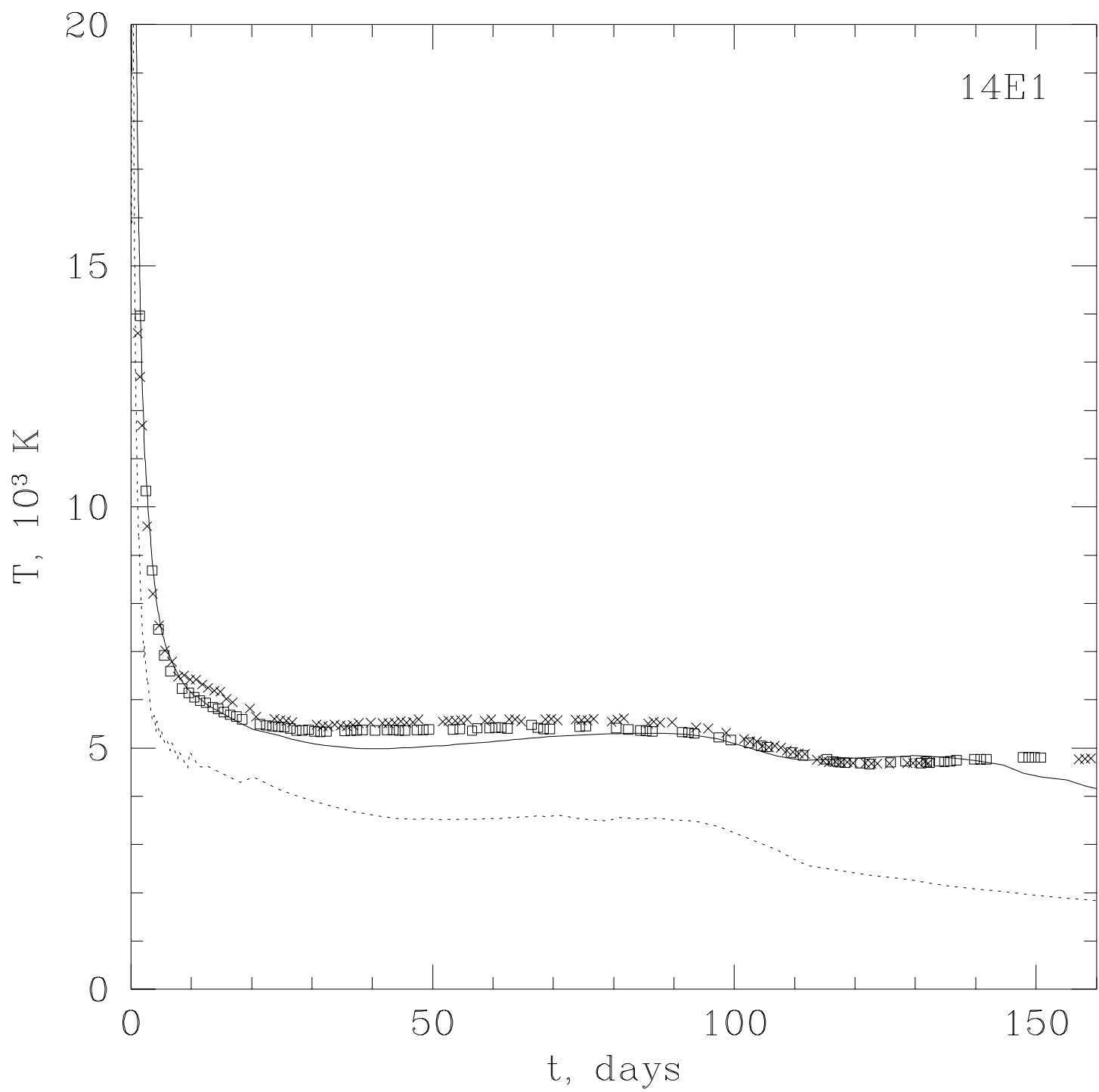
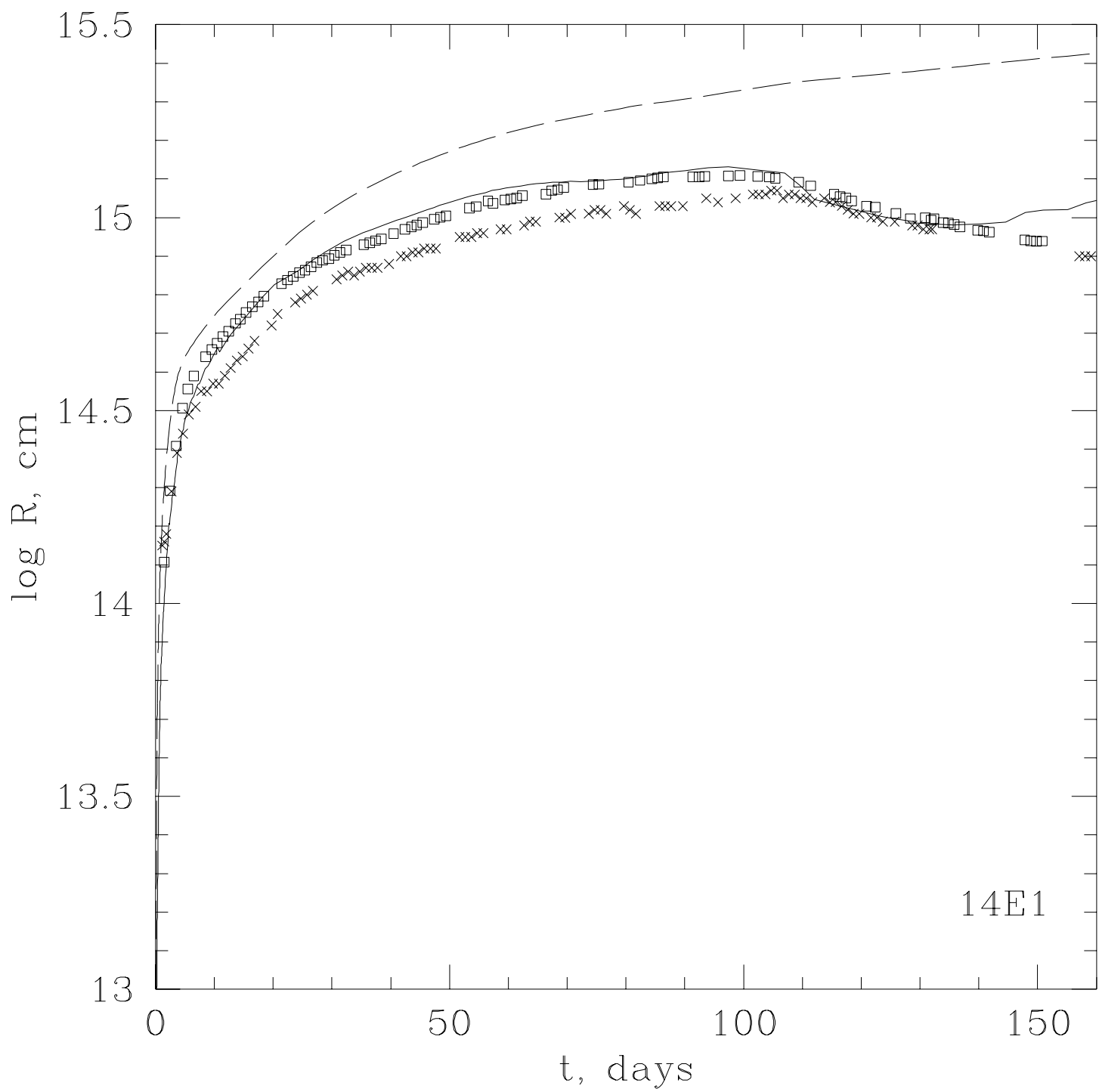
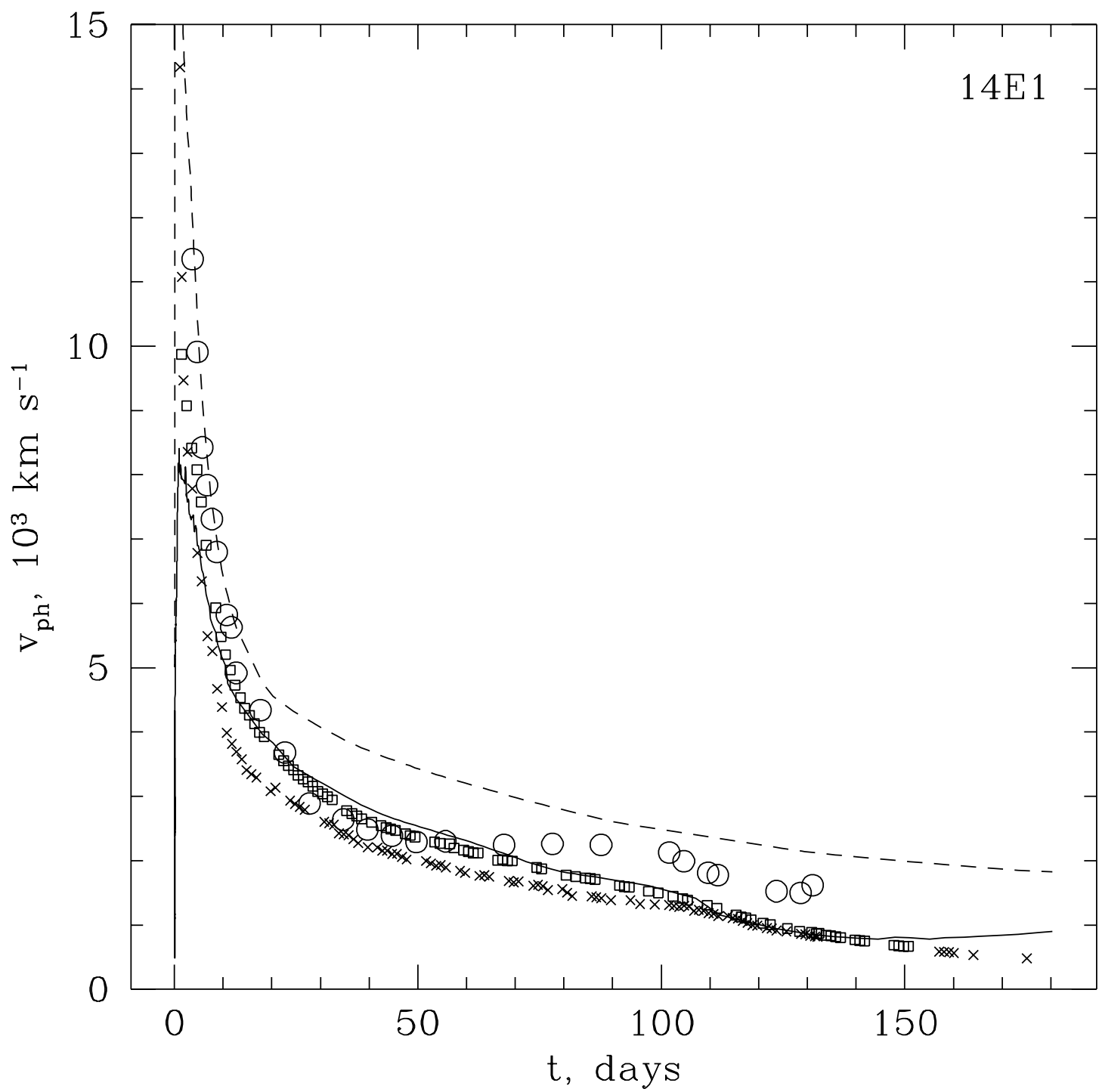
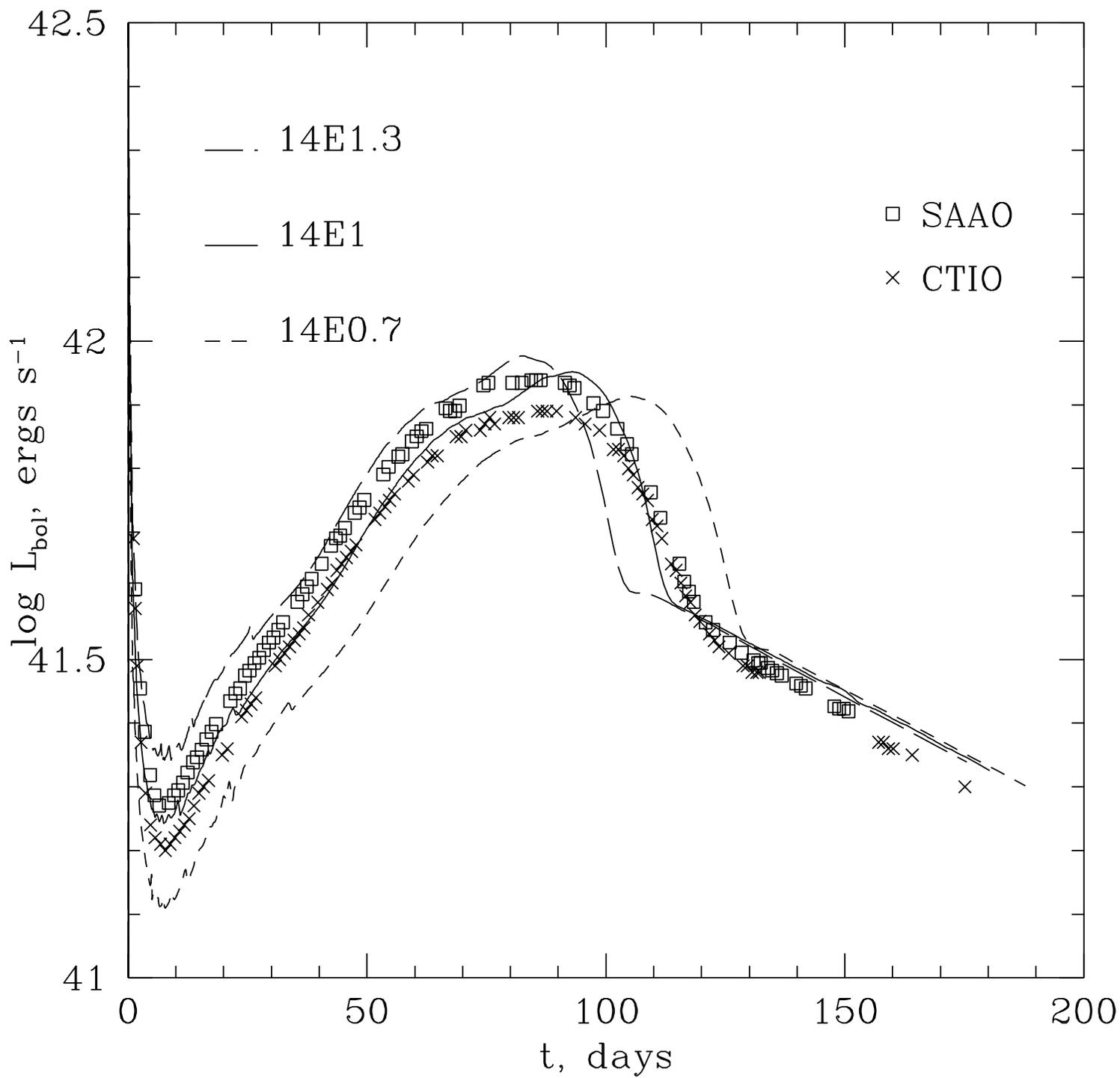
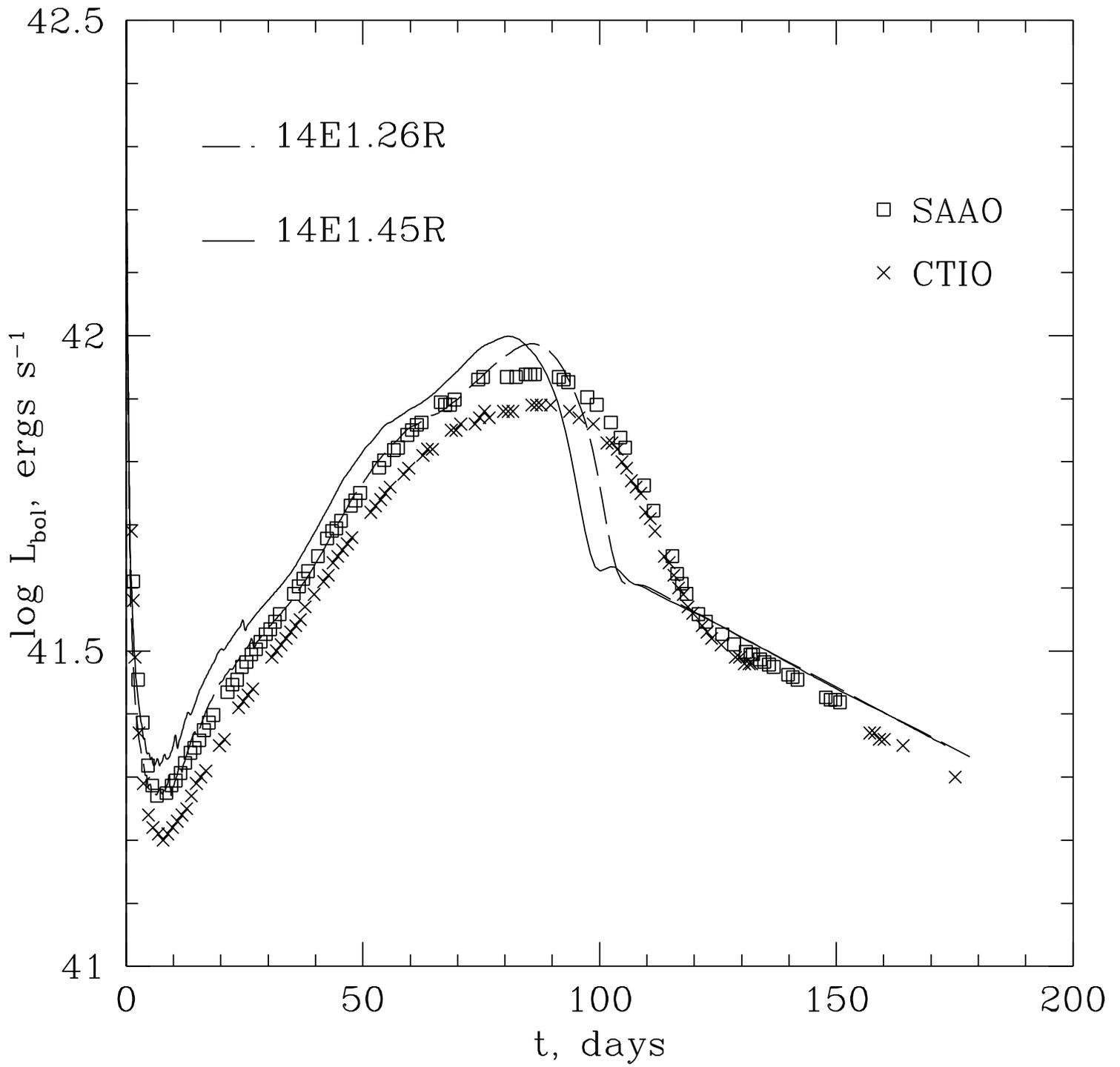
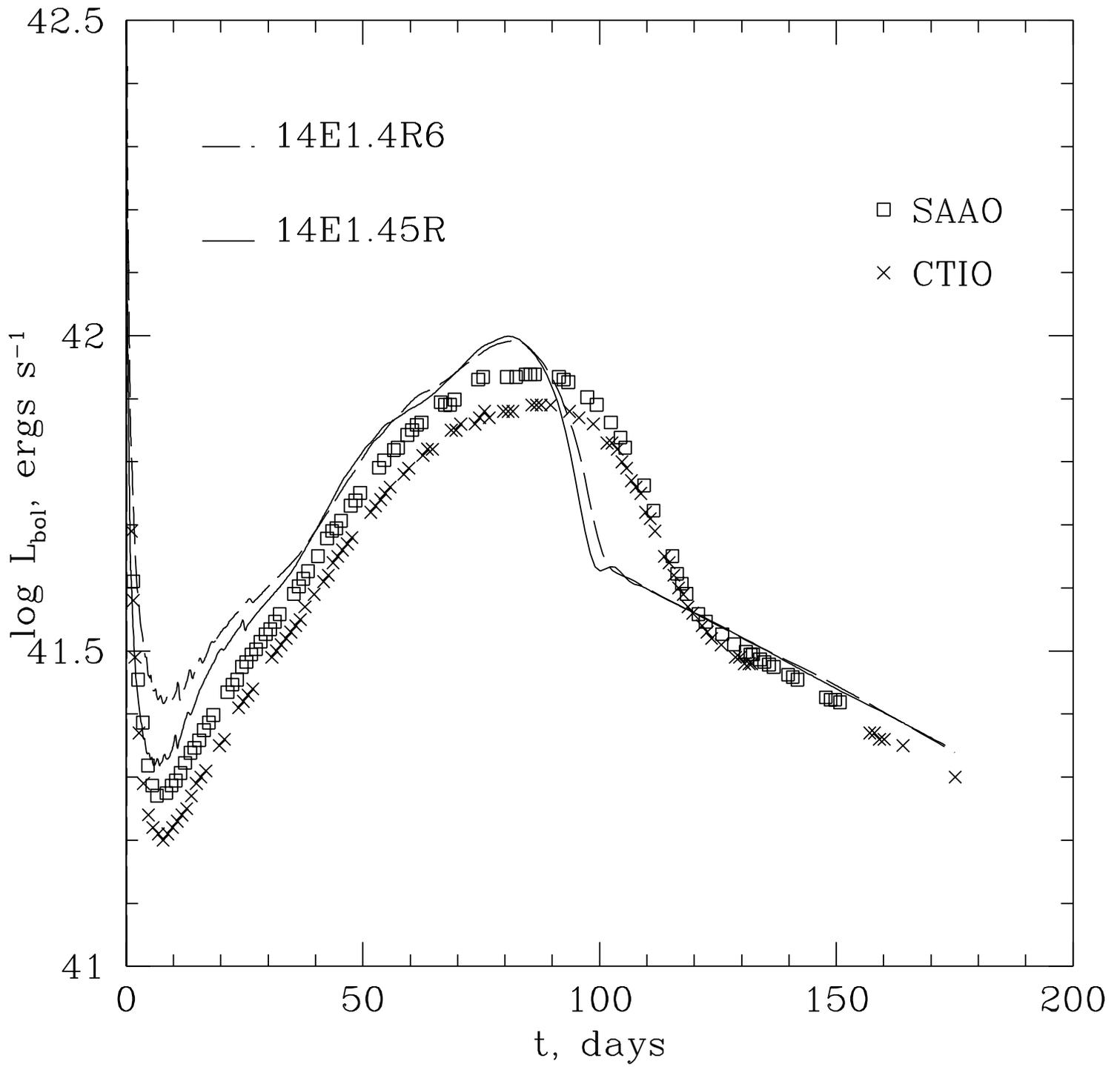
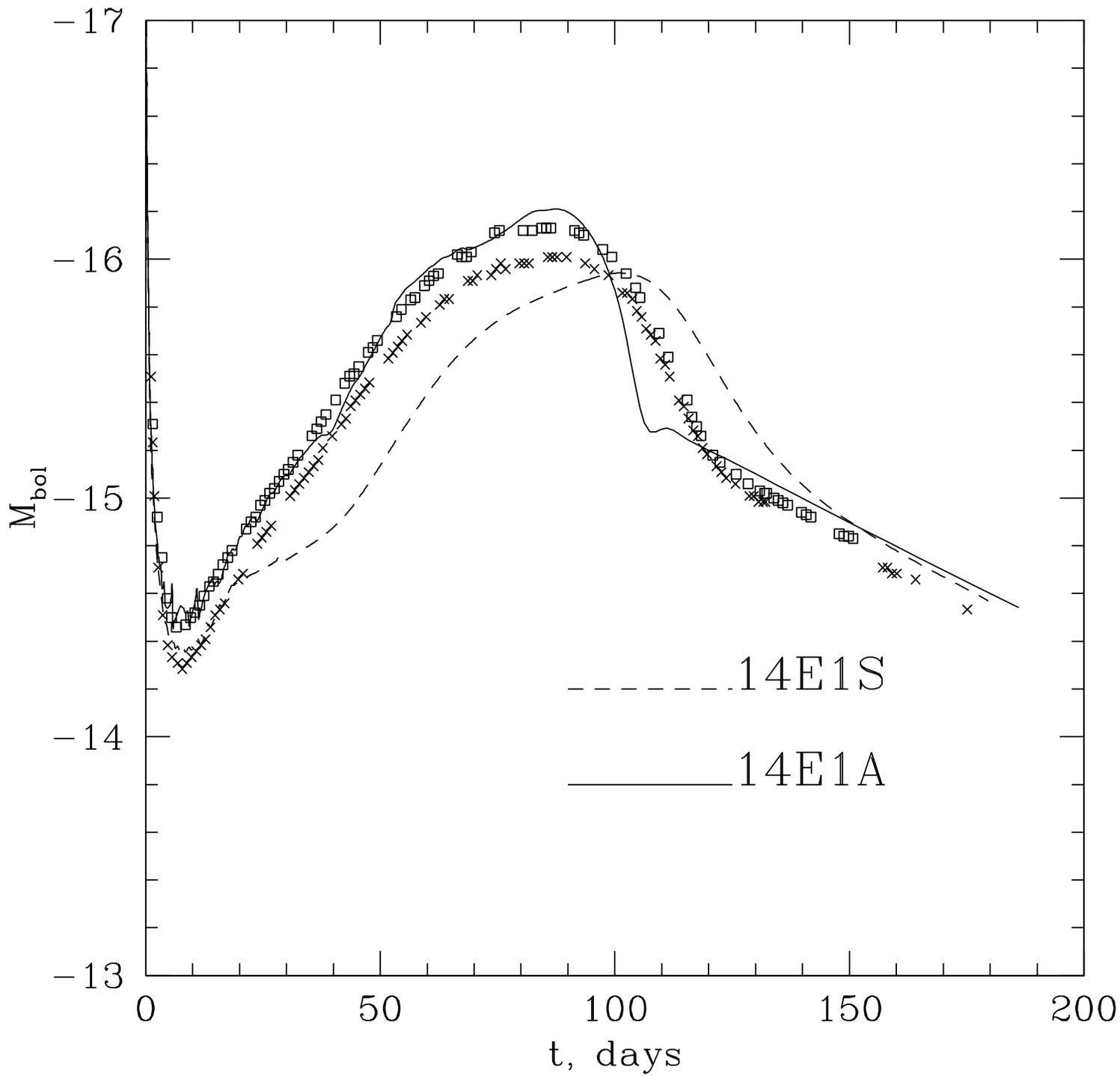
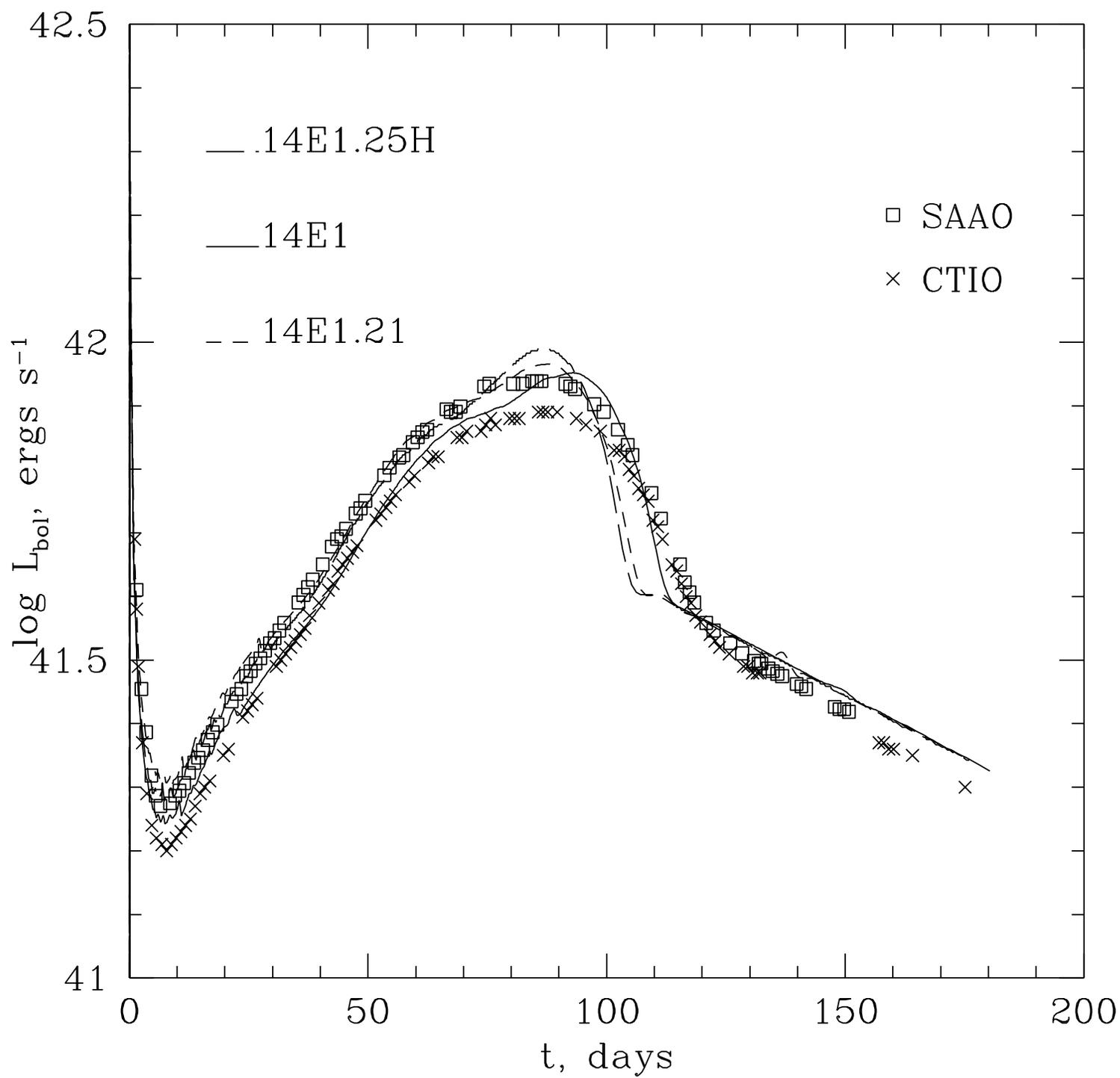
\end{document}